\documentclass[twocolumn]{aastex631}

\usepackage{pbox, cellspace}
\usepackage{hyperref}
\renewcommand{\arraystretch}{2}

\usepackage[T1]{fontenc}
\usepackage{graphicx}	
\usepackage{amsmath}	
\usepackage{amssymb}	
\usepackage{wrapfig}
\usepackage{natbib}
\usepackage{xcolor}
\usepackage{ulem}
\usepackage{soul}
\usepackage{array}
\usepackage{tabularx}
\definecolor{blazeorange}{rgb}{1.0, 0.4, 0.0}
\definecolor{seagreen}{rgb}{0.18, 0.55, 0.34}
\definecolor{rufous}{rgb}{0.66, 0.11, 0.03}
\definecolor{royalfuchsia}{rgb}{0.79, 0.17, 0.57}
\definecolor{scarlet}{rgb}{1.0, 0.13, 0.0}
\definecolor{royalpurple}{rgb}{0.47, 0.32, 0.66}
\definecolor{darkblue}{rgb}{0, 0, 0.66}

\DeclareRobustCommand{\VAN}[3]{#2}
\let\VANthebibliography\thebibliography
\def\thebibliography{\DeclareRobustCommand{\VAN}[3]{##3}\VANthebibliography}

\hypersetup{linkcolor=blue,citecolor=blue,filecolor=cyan,urlcolor=magenta}
\usepackage{newtxtext,newtxmath}
\usepackage{threeparttable}
\usepackage{mathtools}

\def\d{\mathrm{d}}

\newcommand{\lta}{\lower 2pt \hbox{$\, \buildrel {\scriptstyle <}\over {\scriptstyle \sim}\,$}}
\newcommand{\gta}{\lower 2pt \hbox{$\, \buildrel {\scriptstyle >}\over {\scriptstyle \sim}\,$}}

\begin{document}

\title{Correlated Signatures of Plasma Lensing in Fast Radio Bursts}

\correspondingauthor{Paz Beniamini}
\email{pazb@openu.ac.il}

\author[0000-0001-7833-1043]{Paz Beniamini}
\affiliation{Department of Natural Sciences, The Open University of Israel, P.O Box 808, Ra'anana 4353701, Israel}
\affiliation{Astrophysics Research Center of the Open university (ARCO), The Open University of Israel, P.O Box 808, Ra'anana 4353701, Israel}
\affiliation{Department of Physics, The George Washington University, 725 21st Street NW, Washington, DC 20052, USA}

\author{Pawan Kumar}
\affiliation{Department of Astronomy, University of Texas at Austin, Austin, TX 78712, USA}

\begin{abstract}
Plasma lensing has been proposed to explain narrow spectra, multiple burst copies, frequency drifts, interference fringes, and polarization changes in fast radio bursts (FRBs). Because many of these features are not unique to lensing, a convincing identification requires several correlated observables to be reproduced by one lens model. We derive such relations for a one-dimensional Gaussian plasma lens near a fold caustic. We distinguish the phase separation of the merging images from their observable group-delay difference and identify three regimes: resolved burst copies, overlapping images with negligible mutual coherence, and coherent spectral interference. We connect the magnification, image delay, spectral-envelope width, fringe spacing, differential dispersion, time--frequency caustic curvature, and the brightness and ordering of the third image. We derive separate finite-source constraints from large fold magnification and from fringe visibility, and relate the wave-optics magnification scale to the plasma column of the lens.
High-contrast fringes simultaneously constrain the source size and the mutual coherence of the two image fields, which constrains the intrinsic coherence of the FRB radiation once the observed spectrum and instrumental bandpass are accounted for.
We also derive the conditions for differential Faraday rotation in magnetized lenses. Finally, we show that clean lensing by a volume-filling turbulent screen requires fine tuning between rare strong fluctuations and caustic confusion, suggesting that sparse sheets, filaments, interfaces, or shocked clumps are more promising lensing structures.
\end{abstract}


\section{Introduction} 
\label{sec:intro}
Fast radio bursts (FRBs) are millisecond-duration radio transients whose observed properties can be modified substantially during propagation through ionized plasma. In addition to dispersion and multipath scattering, sufficiently compact plasma inhomogeneities can act as refractive lenses. Unlike gravitational lensing, plasma lensing is strongly chromatic: image locations, magnifications, and propagation delays all vary with frequency. A plasma lens can therefore produce narrow spectral amplification features, multiple burst images, non-standard arrival-time tracks, interference fringes, and polarization changes. These signatures can probe both the plasma surrounding the source and the transverse size of the radio-emitting region.

Plasma lensing and scintillation are fundamentally connected propagation mechanisms; both arise from spatial variations in the plasma phase \citep[e.g.][]{Ricket1990,Narayan1992,Feldbrugge2019}. Operationally, we use ``lensing'' for situations in which one identifiable structure produces a small number of correlated images or caustics, and ``scintillation'' when many stochastic propagation paths contribute simultaneously. The transition between these regimes can be smooth, particularly in a strongly turbulent screen.

FRBs are especially sensitive probes of plasma lensing. Their short durations permit small differential delays to be measured, their compact emitting regions can avoid averaging over fine lens structure, and their electric fields can interfere when delayed images retain mutual coherence. This last property offers a probe not supplied by brightness-temperature arguments: those arguments require a coherent emission mechanism \citep{Katz2014,Kumar+17}, but do not determine the first-order temporal coherence function of the electric field. A lens acts as a delayed-copy interferometer, so fringe visibility can constrain the source size and, given the observed spectrum, the temporal coherence of the radiation. Repeating FRBs further allow the chromatic evolution of a lensing event to be followed over time.

There is substantial evidence that other compact radio sources undergo lensing by discrete plasma structures. Extreme scattering events are large, chromatic flux variations commonly attributed to refracting interstellar structures \citep{Fiedler1987,Clegg1998}, whose proposed geometries include compact clouds, filaments, and corrugated sheets \citep[e.g.][]{WalkerWardle1998,PenKing2012,Bannister2016,Tuntsov2016}. Pulsars provide stronger diagnostics: ESEs show correlated changes in dispersion and scintillation \citep{Coles+15}, while arcs and arclets reveal localized, highly anisotropic images \citep{Steinberg2001,Walker2004,Cordes2006,Brisken2010}. Such structures appear common and are associated with SNRs, H\,\textsc{ii} regions, bubbles, and bow shocks \citep{Stinebring2022,Main2023B,Ocker2024}. In PSR~B1957+20, plasma lensing differentially magnified pulse components and resolved emission regions separated by tens of kilometres \citep{Main+2018}.

Plasma lensing has been invoked to explain various FRB properties  \citep{Cordes2017,ErRogers2018,GrilloCordes2018,Er2020}. Host-galaxy lenses may produce large magnifications, narrow and time-variable spectral peaks, multiple burst copies with different apparent dispersion measures, and interference when the image delays are small enough. The complex spectral structure of FRB~20121102A has been discussed in this context \citep{Gajjar2018,Hessels+19,Platts2021,Lekov2022}. Plasma-lens models have also been proposed to explain upward and downward frequency drifts, chromatic activity windows, depolarization, apparent circular polarization, and abrupt polarization-angle transitions; for example, the frequency-dependent activity window of FRB~20180916B \citep{Pastor-Marazuela2021,Bethapudi2023} has been interpreted as a possible caustic-magnification effect \citep{LiLensing2026}. 
These examples demonstrate that plasma lensing might be an important aspect of FRB phenomenology and separating lensing effects from intrinsic ones will be key to deciphering the underlying intrinsic FRB physics.

Establishing that lensing is responsible for a given observation is difficult. Many proposed signatures are not unique. Intrinsic sub-burst drift can resemble a chromatic image track; intrinsically narrow emission can resemble a lensing magnification envelope; ordinary scintillation can produce spectral modulation; and bursts from a highly active repeater may occasionally have similar time--frequency morphologies by chance. Isolated features should therefore be interpreted with caution \citep{Platts2021}. This difficulty is illustrated by the recent analysis of FRB~20240114A, in which burst-rate enhancements, predicted magnification and demagnification intervals, burst bandwidths, energies, and proposed ``carbon-copy'' pairs \citep{Uttarkar2026} could not be described consistently by a single Gaussian lens \citep{Niu2026}.

Plasma lensing and scintillation can, in principle, transform an intrinsically broadband emission into the narrow spectra often observed in FRBs (e.g. \citealt{Yang2023,Kumar2024}). Very narrow spectral widths are common, particularly among bursts from repeating sources \citep{CHIME_1st_cat,Pleunis2021B}. However for these widths to be generated by lensing, the source must occupy a small part of the lensing cross-section and the observing band must intersect the magnification peak. This makes it statistically unlikely for propagation-induced spectral narrowing to explain the prevalence of narrow-band FRBs \citep{Kumar2024}.

We therefore set out to examine whether plasma lensing predicts a robust set of observables that distinguishes propagation from intrinsic emission structure. The delay between two images determines either the separation of resolved burst copies or, in the coherent regime, the spectral fringe spacing. The same image geometry fixes their relative magnifications, differential plasma columns, and time--frequency tracks. The smooth caustic magnification envelope is distinct from the much finer interference-fringe scale. A third image may precede or follow the bright near-caustic pair, depending on the caustic involved. If the lens is magnetized, the ray separation also determines the differential rotation measure and associated polarization effects \citep{BKN2022,LiLensing2026}. Finally, both the magnification and the survival of coherent fringes constrain the source size.

This use of multiple images as probes of coherence and intervening plasma is related to our study of gravitationally lensed FRBs propagating through a plasma screen \citep{KB2023}. There, plasma scattering modifies gravitational-image magnifications, delays, and polarization. Here the plasma itself creates the images, linking these effects through one chromatic lens geometry.

In this work we derive robust relations for a one-dimensional Gaussian plasma lens. We connect magnification, image delays, spectral envelopes, interference fringes, differential dispersion, polarization effects, and source-size limits, and identify combinations of observables that can constrain the lens size, distance, column density, and magnetic field. We also examine whether isolated strong lenses can arise in volume-filling turbulent screens, and when such screens instead produce a confused network of caustics or ordinary scintillation. Readers primarily interested in observational diagnostics and search strategies may proceed directly to \S \ref{sec:Observe}, where the principal correlated signatures are summarized and their practical implications are discussed.

\section{Plasma lensing: overview}
\label{sec:plasmalens}

\begin{figure}
\centering
\includegraphics[scale=0.08]{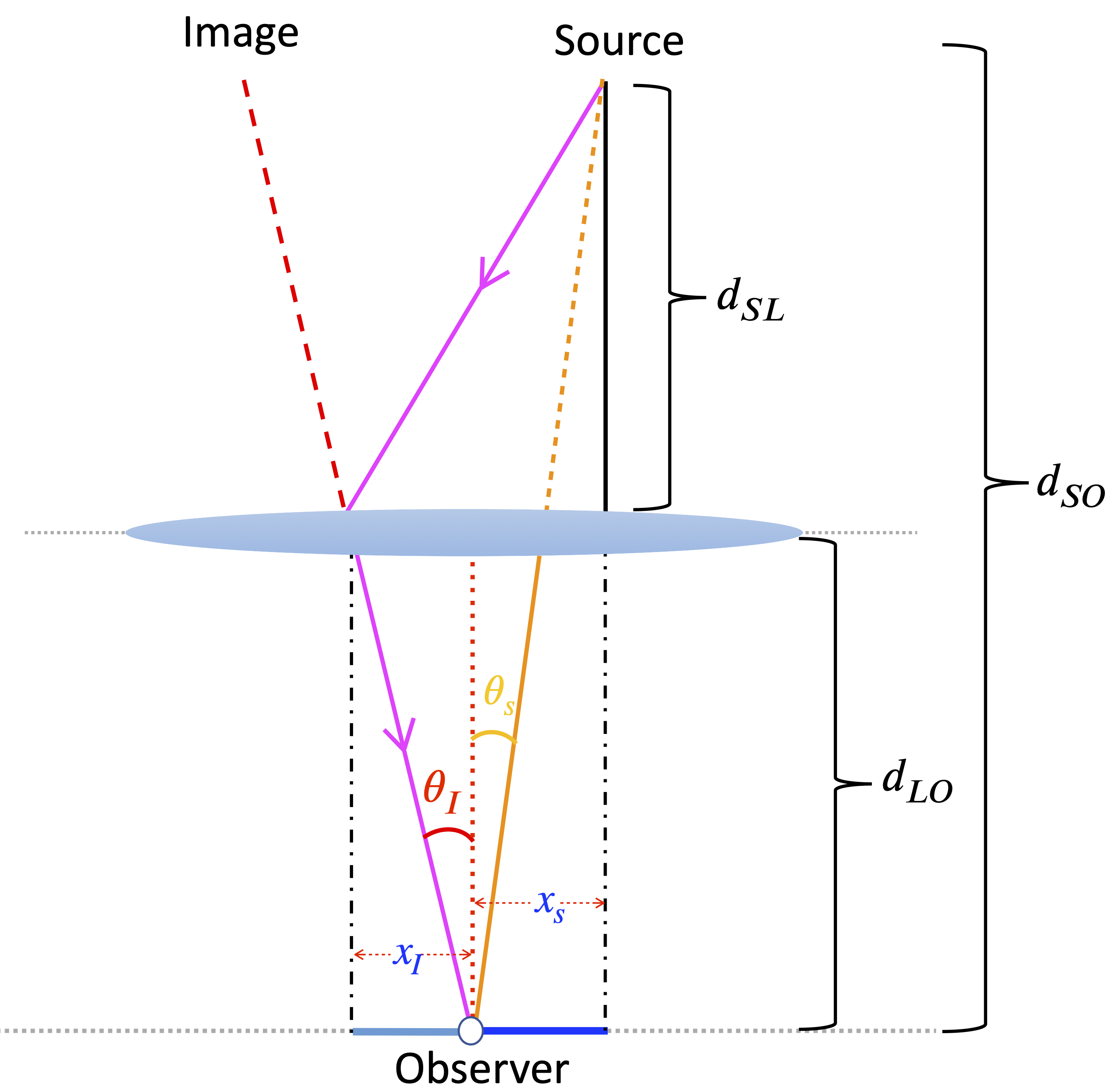}
\caption{Schematic illustration of plasma lens geometry. The ray trajectory from the source to the observer through the plasma lens, together with the angular positions of the source and its image, and the geometric parameters used in the lensing calculations.
}
\par\vspace{1em}
\label{fig:plasma-lens-schematic}
\end{figure}

We first present the geometric setup and general properties of plasma lensing.
The wave amplitude at the observer location when there is a plasma screen between the source and the observer is given by the Fresnel-Kirchhoff integral over the plasma screen surface, e.g. \cite{Feldbrugge2019,jow2021,BKN2022}:
\begin{equation}
\label{eq:Ai}
A_i(\nu) \propto \int d^2{\vec \theta} \exp\left[ \frac{ i \pi (\vec\theta-\vec\theta_s)^2}{\theta_F^2}+ i\Phi_p(\vec\theta,\omega)\right],
\end{equation}
where $\vec\theta_s$ is the angular location of the source in the sky relative to observer-lens line, $\vec\theta$ is
the angular position of a point in the plasma lens plane,
\begin{equation}\label{fresnel}
\theta_F=\left(\frac{\lambda d_{\rm SL}}{d_{\rm LO}d_{\rm SO}}\right)^{1/2},
\end{equation}
is the Fresnel angle, $d_{\rm SL}$ is the distance from the FRB source to the lens, $d_{\rm LO}$ from the lens to the observer, and $d_{\rm SO}$ from the FRB source to the observer; see Fig. \ref{fig:plasma-lens-schematic} for the lens geometry and the notation.
The Fresnel angle corresponds to an additional geometric phase of $\pi$ relative to the undeflected path. The remaining phase, $\Phi_p$, is acquired in crossing the plasma screen at the angular location $\vec\theta$ and, for a frequency much higher than the plasma frequency, is given by
\begin{equation}
\Phi_p(\theta,\omega)\!=\!-\omega\!\int \!\frac{dz}{c}\frac{2\pi e^2 n_e(\vec\theta, z)}{m_e\omega^2} \equiv -\frac{\xi N_e(\vec\theta)}{\omega} \!=\! -\lambda\, r_e N_e(\vec\theta)
\end{equation}
where $r_e\equiv e^2/m_e c^2$ is the classical radius of an electron, $z$ is the coordinate perpendicular to the plasma lens plane,
\begin{equation}
\xi\equiv\frac{2\pi e^2}{m_e c}=0.053\,{\rm cm^2\,s^{-1}} \ {\rm and} \  N_e(\vec\theta)\equiv\int dz \,n_e(\vec\theta,z)
\end{equation}
is the electron column density of the lens at the angular position $\vec\theta$ ($n_e(\vec\theta,z)$ being the volumetric number density).
The image's angular location is given by the extremum of the phase function
\begin{equation}\label{phase-fun}
\Phi \equiv \frac{\pi (\vec\theta - \vec\theta_s)^2}{\theta_F^2}+\Phi_p(\vec\theta,\omega),
\end{equation}
where the first term is the geometric phase. Explicitly,
\begin{equation}
 \label{image-loc}
\frac{\partial \Phi}{\partial \theta_i} \equiv \partial_i \Phi=0 \quad {\rm or} \quad
\frac{2\pi(\theta_{I i} - \theta_{si})}{\theta_F^2}-\omega^{-1} \xi \partial_i N_e(\vec\theta_I)=0. 
\end{equation}
Eq. \ref{image-loc} shows that the angular position of the image in the sky is frequency dependent.

The amplification of flux due to the lens is given by
\begin{equation}
\mu = \frac{4 \pi^2}{ \theta_F^4 {\rm det}\left[ \partial_i \partial_j \Phi \right] }.
\label{mag}
\end{equation}
$\mu$ denotes the signed geometric-optics magnification while the observable flux magnification of an isolated image is $|\mu|$.
We use Eq. \ref{phase-fun} to write the $2\times2$ `amplification' matrix elements
\begin{equation}
\partial_i \partial_j \Phi = 2\pi \theta_F^{-2} \delta_{ij} - \omega^{-1} \xi \partial_i\partial_j N_e(\vec\theta),
\label{mag-mat}
\end{equation}
where $\delta_{ij}$ is the standard Kronecker delta function.

Since $\partial_i \partial_j \Phi $ is a $2\times2$ symmetric matrix, it can be diagonalized by a coordinate rotation, giving:
\begin{gather}
\partial_i\partial_j N_e(\vec\theta) =
\begin{pmatrix}
A_1  & 0 \\
0   & A_2
\end{pmatrix}
 \quad {\rm \&} \quad  \partial_i \partial_j \Phi = {2\pi\over \theta_F^2}
\begin{pmatrix}
B_1^{-2}  &  0 \\
0  &  B_2^{-2}
\end{pmatrix}
\label{mag-mat1}
\end{gather}
where
\begin{equation}
  B_k \equiv {1\over \sqrt{1 - \psi A_k}}, \; \psi \equiv {\theta_F^2 \xi\over 2\pi \omega} = \psi_0 \left( {\omega_0\over\omega}\right)^2, \;
      \psi_0 \equiv {\theta_F^2(\omega_0)\xi\over 2\pi\omega_0}, 
\end{equation}
$\omega_0$ is a characteristic frequency, often chosen near the frequency at which the eigenvalue approaches criticality for the image under consideration and $\psi_0$ is independent of wave frequency. The shape of the region in the lens plane from which all waves arrive at the observer with a relative phase difference of $<\pi$, such that their amplitudes add constructively, can be approximated as an ellipse when cubic and higher-order terms in angle are neglected. For a circular lens, $B_1=B_2$, and for a highly elongated lens either $B_1 \gg B_2\sim 1$ or $B_2 \gg B_1\sim 1$. The magnification of the lens follows from Eqns. \ref{mag} \& \ref{mag-mat1}\footnote{When one of the eigenvalues flips sign, $B_k$ should be understood through $B_k^{-2}=1-\psi A_k$. The magnification is obtained by taking the absolute value of the signed expression.},
\begin{equation}
      \mu = B_1^2 B_2^2.
  \label{mag2}
\end{equation}
The magnification diverges along the critical curves in the lens plane, which are given by either $B_1^{-1}=0$ or $B_2^{-1}=0$. The mapping of a critical curve into the source plane is known as a caustic curve. When a source crosses a caustic, the image multiplicity changes by two: either two images merge and disappear, or a new pair of images is created, depending on the side of the caustic on which the source lies.

Strong amplification requires cancellation of the geometric and plasma phases over an area larger than $\theta_F^2$. Expanding the phase quadratically about an image gives coherent dimensions $\theta_F B_1$ and $\theta_F B_2$, and hence $\mu=B_1^2B_2^2$. The required column variation is set by the curvature of $N_e$, rather than its absolute value. Therefore, adding a uniform plasma column changes the common dispersive delay but not the image locations or magnifications.
We consider next the special case of a one-dimensional Gaussian plasma lens in some detail. 

\section{Gaussian lens in 1D}
\label{sec:Gausslens}

\subsection{Lens strength and formation of caustics}
\label{sec:gauss_preliminaries}
A one-dimensional Gaussian lens is described by a Gaussian column-density profile along the x-axis, with no dependence on the transverse coordinate (y):
\begin{equation}
 N_e(\theta_{x}, \theta_y) = N_0\exp{-[\theta_{x}^2 d_{\rm LO}^2/a^2]} 
  \label{gaussian-lens-prof}
\end{equation}
where $\theta_{x}$ is the angular position of a point in the lens plane along the $x$-direction, and $a$ is the length in the lens-plane over which the plasma density decreases by a factor $e$. Using Eq. \ref{image-loc} we obtain the angular position of images, 
\begin{eqnarray}
\label{eq:ImageLOCGauss}
& \theta_{Ix}-\theta_{sx}=-\frac{\lambda r_e \theta_F^2\theta_{Ix} d_{\rm LO}^2}{\pi a^2} N_0 e^{-\theta_{Ix}^2 d_{\rm LO}^2/a^2} \\ &
=-\frac{\lambda r_e R_F^2\theta_{Ix} }{\pi a^2} N_0 e^{-\theta_{Ix}^2 d_{\rm LO}^2/a^2} \quad ; \quad \theta_{Iy}=\theta_{sy}, \nonumber
\end{eqnarray}
where $(\theta_{Ix},\theta_{Iy})$ is the image location in the $x$ and $y$ directions, $(\theta_{sx}, \theta_{sy})$ is the source location, and $R_{\rm F}\equiv \sqrt{{\lambda d_{\rm SL} d_{\rm LO}\over d_{\rm SO} }}=\theta_{\rm F}d_{\rm LO}$
is the Fresnel length.
We define a dimensionless parameter that determines the lens strength
\begin{equation}
\label{eq:alpha}
    \alpha \equiv \frac{\lambda r_e R_F^2N_0}{\pi a^2}\approx 24\, \nu_9^{-2}d_{\rm SL,19}\mbox{DM}_{0} a_{13}^{-2}
\end{equation}
where $\mbox{DM}_0$ is the dispersion measure for a sightline going through the center of the lens (same as $N_0$ but measured in $\mbox{pc cm}^{-3}$ as standard in radio astronomy) and the rightmost side of Eq. \ref{eq:alpha} is for a host galaxy lens.
Defining the typical geometric delay scale associated with a transverse lens size $a$ as $T_G=\Phi_G/\omega=\pi a^2/(\omega R_F^2)$, and the magnitude of the plasma phase-delay as $T_p^{\rm abs}=|\Phi_0|/\omega$ (where $\Phi_0$ is the plasma phase accumulated through the screen center, $\Phi_{\rm p}(0)$), we find $\alpha=T_p^{\rm abs}/T_G$.
In the strong-lensing regime (\(\alpha \gg 1\)), the plasma phase delay dominates over the geometric component.

Expanding the Gaussian near its axis gives
\begin{equation}
 f_{\rm foc}\simeq-\frac{\pi a^2}{\lambda^2r_eN_0},
\end{equation}
where $f_{\rm foc}$ is the focal distance, at which initially parallel rays would be brought by the lens to a focus (negative for a diverging lens). Hence
\begin{equation}
 \alpha={\lambda^2 r_e N_0\over \pi a^2}
   {d_{\rm SL}d_{\rm LO}\over d_{\rm SO}}
   =\frac{d_{\rm eff}}{|f_{\rm foc}|},\qquad
 d_{\rm eff}=\frac{d_{\rm SL}d_{\rm LO}}{d_{\rm SO}}.
\end{equation}
Strong lensing requires a focal length shorter than the effective propagation distance \footnote{More precisely, the condition becomes $|f_{\rm foc}|\lesssim d_{\rm eff}/\alpha_{\rm cr}$, or equivalently $\alpha\gtrsim \alpha_{\rm cr}$ with $\alpha_{\rm cr}=0.5 e^{3/2}$ as shown explicitly below.}.

It is convenient to work with the following dimensionless parameters for the source and image locations:
\begin{eqnarray}
 & u_{I}\equiv \theta_{Ix}d_{\rm LO}/a, \quad u_{s}= {\theta_{sx}d_{\rm LO}\over a}\approx \frac{d_{\rm LO}}{a\,d_{\rm SO}}[\vec{x}_s+\frac{d_{\rm SL}}{d_{\rm LO}}\vec{x}_o]
\end{eqnarray}
Without loss of generality, we take the observer location $\vec{x}_o=0$. For a host galaxy lens, we then get $u_{s} \approx \frac{x_{\rm sx}}{a}$.

The equation for image location can be written in a simplified form in terms of $\alpha$ and $u_{s}$
\begin{equation}
\label{eq:uIx}
 u_{I}(1+\alpha \exp{[-u_{I}^2]})= u_{s},
\end{equation}
and the magnification of the image is given by
\begin{equation}
\label{eq:muGausslens}
\mu=\bigg[1+\alpha\bigg(1-2u_{I}^2\bigg)\exp{\{-u_{I}^2\}}\bigg]^{-1}.
\end{equation}
The magnification equation given above is the same as Eq. \ref{mag2} applied to a 1D lens with the column density profile given in Eq. \ref{gaussian-lens-prof}; in the notation of \S\ref{sec:plasmalens}, the nonzero Hessian eigenvalue is $A_x$ with 
$\psi A_x = \alpha (2u_I^2-1)\exp(-u_I^2)$ while $A_y = 0$. Thus $B_y=1$ and Eq. \ref{mag2} reduces to Eq. \ref{eq:muGausslens} for 1D Gaussian lens.

From Eq. \ref{eq:muGausslens}, $\mu^{-1}$ crosses zero if the minimum of 
\begin{equation}
  f(u_{I})\equiv \alpha(1-2u_{I}^2)\exp[-u_{I}^2]<-1. 
  \label{lens-def-f}
\end{equation}

The minimum point of $f(u_{I})$ occurs at $u_{I}=\sqrt{3/2}$ and its value is $f(\sqrt{3/2})=-2\alpha e^{-3/2}$. Therefore, critical curves (curves in the image plane where $\mu\to \infty$) exist for $\alpha>0.5 e^{3/2}\equiv \alpha_{\rm cr}$, and their locations are determined by solutions of the equation $f(u_{I*}) = -1$.
When they exist, critical curves occur in pairs. The locations of the two critical curves in the image plane, for $\alpha\gg 1$, are approximately given by
\begin{equation}
\label{eq:caustic-Iplane}
\begin{aligned}
u_{IA} &\approx \frac{1}{\sqrt{2}} + {e^{1/2}\over \alpha \sqrt{8}}, \\[5pt]
u_{IB} &\approx \sqrt{\log\!\left(2\alpha \log \alpha\right)}\equiv \sqrt{\mathcal{L}_{\alpha}}.
\end{aligned}
\end{equation}
where for brevity, $\mathcal{L}_{\alpha}$ is defined as above as it appears in many of the expressions below ($\mathcal{L}_{\alpha}=u_{IB}^2$ is accurate to $\sim 10\%$ for $\alpha\gtrsim 10$).

For $\alpha \gg 1$, the locations in the source plane corresponding to critical curves A \& B in the image plane, known as caustics, are obtained by substituting Eq. \ref{eq:caustic-Iplane} into Eq. \ref{eq:uIx}
\begin{eqnarray}
\label{eq:caustic-Splane}
&  u_{sA} \approx \frac{\alpha}{\sqrt{2e}} + \frac{1}{\sqrt{2}} + \rm O(\alpha^{-1}),\\ &
  u_{sB} \approx
\sqrt{\mathcal{L}_{\alpha}}\bigg[1+\frac{1}{2\mathcal{L}_{\alpha}-1}\bigg]\sim \mbox{few}.\nonumber
\end{eqnarray}

Eq. \ref{eq:uIx} has either 1 or 3 images for a point source depending on $\alpha$ \& $u_{s}$. 3 solutions only exist when the LHS of Eq. \ref{eq:uIx} has a local maximum and a local minimum. This is equivalent to the condition that caustics exist, i.e. $\alpha>\alpha_{\rm cr}$. Multiple images then form when the source lies between $u_{sB}$ and $u_{sA}$.

\begin{figure*}
\centering
\includegraphics[width=1.\textwidth]{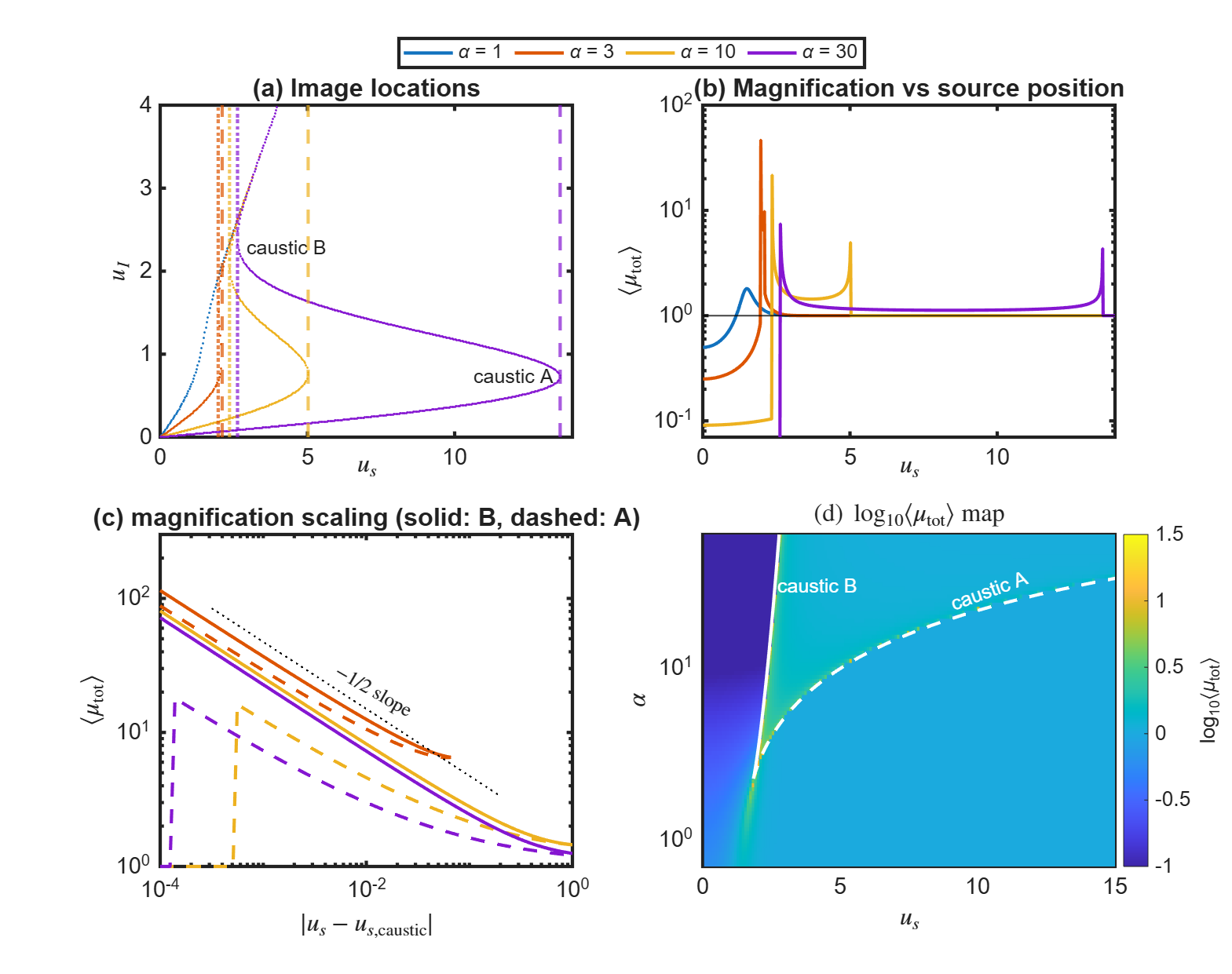}
\caption{Properties of 1D Gaussian lenses in geometrical optics. Top left: Image locations vs. source positions (Eq. \ref{eq:uIx}). Vertical lines denote source locations that correspond to caustics (leading to $\mu\to \infty$ for $\alpha>\alpha_{\rm cr}$). Top right: Corresponding geometric-optics total magnification, obtained by summing $|\mu|$ from Eq. \ref{eq:muGausslens} over all images (1 or 3 depending on $\alpha$ and $u_s$). Note that for a given DM$_{0}$ and $\nu$ the achievable, wave-optics limited, magnification is capped by Eq. \ref{eq:muw}.
Bottom left: scaling of total magnification with source position near caustics. As in panel B, the results follow Eq. \ref{eq:muGausslens} directly and do not assume a near-caustic approximation. At large $|u_s-u_{s,\rm caustic}|$, the sum of magnification from the images is saturated at 1.} Bottom right: heat-map showing magnification as a function of $\alpha$ and $u_{s}$. 
\label{fig:imageandmagnification}
\end{figure*}

Within the caustic interval the three solutions are approximately\footnote{While these approximations are useful, we caution that they should generally not be plugged back into the magnification relation (Eq. \ref{eq:muGausslens}). This is because near caustics, where magnification is largest, it changes extremely rapidly over very small changes in $u_{I}$. An approximation for $u_{I}$ near caustics is provided by Eq. \ref{eq:caustic-Iplane}.} an inner image, $u_I\simeq u_s/(1+\alpha)$ with $|\mu|\simeq(1+\alpha)^{-1}$; an intermediate image with $u_I^2\simeq\log(\alpha/u_s)+\log u_I$; and an outer image with $u_I\simeq u_s$ and $|\mu|\simeq1$. 
For $u_{s}<u_{sB}$ only the inner solution survives, while for $u_{s}>u_{sA}$ only the outer solution does. Large magnification occurs only at the threshold when two of the images (inner and intermediate or intermediate and outer) almost coincide. Figure \ref{fig:imageandmagnification} shows the numerical solutions of Equations \ref{eq:uIx}, \ref{eq:muGausslens}.

\begin{figure}
\centering
\includegraphics[width=0.43\textwidth]{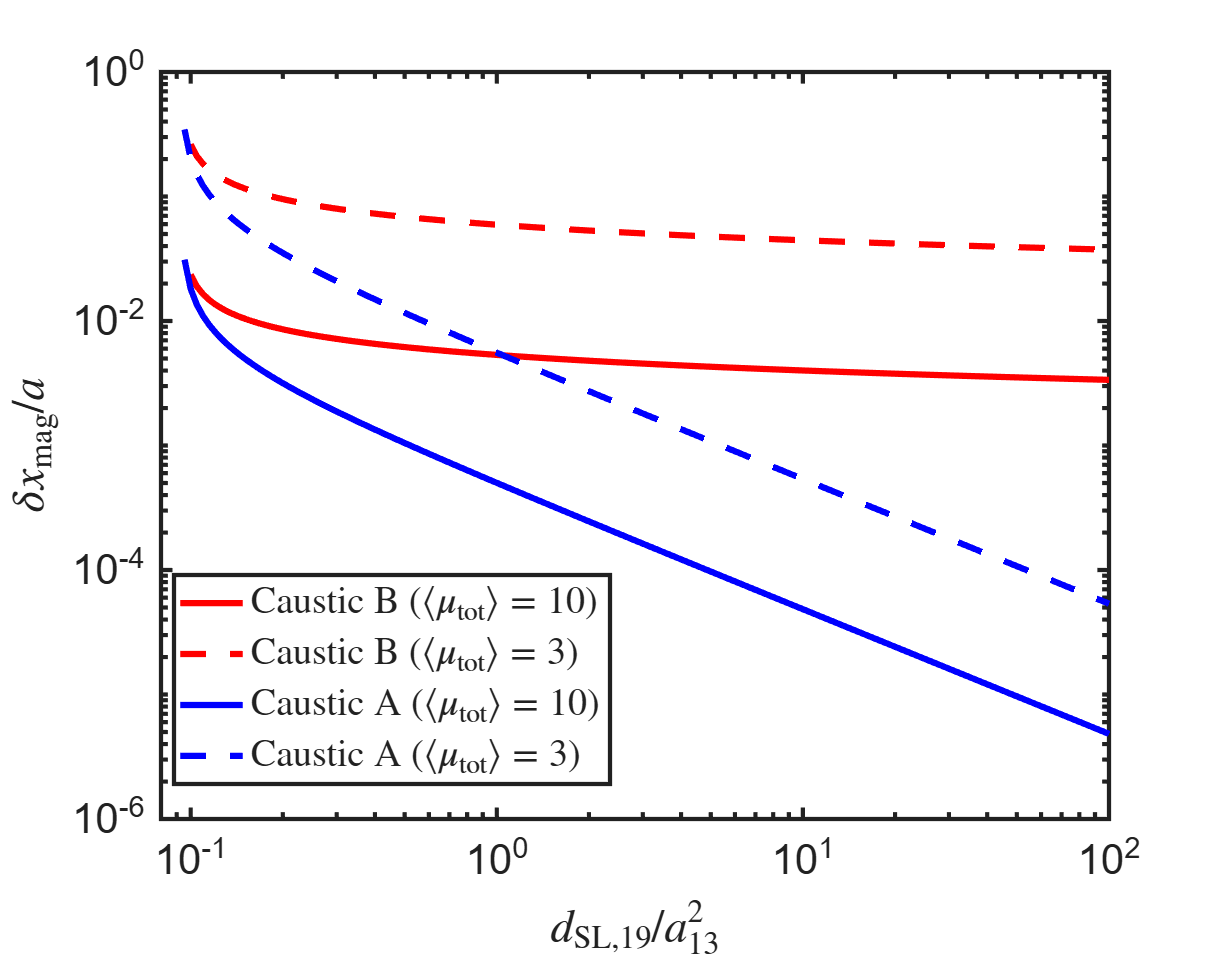}
\caption{Maximum lateral source size (normalized by lens size) for a magnification due to plasma lensing by factors of $\langle \mu_{\rm tot}\rangle=10$ and $\langle \mu_{\rm tot}\rangle=3$ near caustics A and B. We have taken here: $\mbox{DM}_0=1\mbox{pc cm}^{-3}$ and $\nu=1$\,GHz. Note that $\alpha$ is linear in $d_{\rm SL}/a^2$: $\alpha$ varies here from $\alpha_{\rm cr}$ to $\alpha=2.4\times 10^3$.} The maximum source size decreases rapidly with magnification, as $\langle \mu_{\rm tot}\rangle^{-2}$. 
\label{fig:maxsize}
\end{figure}

\subsection{Source size from magnification}

To determine the behavior of the magnification near the critical curves, we consider
$u_{I} = u_{I*} + \epsilon$, where $\epsilon$ denotes a small displacement from the critical curve in the image plane, and $u_{I*}$ is defined by the critical curve condition
$f(u_{I*}) = -1$, with $f$ given in Eq.~\ref{lens-def-f}. Expanding as a series in $\epsilon$,
\begin{eqnarray}
   \label{caustic2}
   &\left.{df\over du}\right|_{u_{I*}}= \frac{2u_{I*}(3-2u_{I*}^2)}{1-2u_{I*}^2}\equiv \eta \implies\nonumber \\ &  f(u_I)\approx -1+\eta \epsilon \quad \&\quad \mu(\epsilon)\approx {1\over \eta \epsilon}. 
  \end{eqnarray}
The new source location, when the image shifts by $\epsilon$ from a critical curve, is obtained by substituting $u_{I}$ into Eq. \ref{eq:uIx},
\begin{equation}
\label{eq:u_s}
  u_{s}\approx -\frac{2u_{I*}^3}{1-2u_{I*}^2}+\frac{\eta \epsilon^2}{2}\approx -\frac{2u_{I*}^3}{1-2u_{I*}^2}+\frac{1}{2\eta \mu^2}.
\end{equation}
A $\delta u_{s}$ shift in the source position leads to a magnification change by $|\delta \mu |= |\eta \mu^3 \, \delta u_{s}|$. Our goal is to estimate how far the source can extend transversely while keeping the flux magnification comparable to its local value, $\mu$. An order of magnitude estimate of this criterion is given by taking
$\delta \mu \approx \mu/2$.
For a source near a caustic, two images are formed with almost equal brightness and are separated by a tiny angular distance for plasma lenses of interest in FRBs (this results in a very short time separation, often much shorter than the FRB duration). Therefore, it is appropriate to add the flux from them, leading to the total observable magnification
$\langle \mu_{\rm tot} \rangle = |\mu_+| + |\mu_-| \approx 2|\mu|$.
Hereafter $\langle\mu_{\rm tot}\rangle$ denotes the smooth, incoherent magnification envelope of the pair \footnote{In the coherent regime the instantaneous magnification oscillates about this value, see \S \ref{sec:images_interact}}. 
The maximum source size, perpendicular to the caustic, that still allows a magnification of order $\langle \mu_{\rm tot} \rangle$ is thus
\begin{equation}
\label{eq:delta_umax}
|\delta u_{\rm mag}| \approx {2\over |\eta|\,\langle \mu_{\rm tot} \rangle^2 }
\end{equation} 
The $\langle\mu_{\rm tot}\rangle^{-2}$ dependence reflects the inverse of the point-source fold-caustic scaling: $\langle\mu_{\rm tot}\rangle\propto |u_s-u_{s,\rm caustic}|^{-1/2}$.
For caustics A and B, the corresponding values of $\eta$, denoted by $\eta_A$ and $\eta_B$, are obtained by substituting Eq.~\ref{eq:caustic-Iplane} into Eq.~\ref{caustic2}. In the $\alpha \gg 1$ limit,
\begin{equation}
   \eta_A \approx -{2\sqrt{2}\,\alpha\over \sqrt{e}}, \quad \eta_B \approx 2\sqrt{\mathcal{L}_{\alpha}}.
    \label{eta-AB}
\end{equation}

For a lens in the host galaxy, $\delta x_{sx} \approx a\,\delta u_{s}$, where $a$ (given in Eq.~\ref{gaussian-lens-prof}) is the characteristic scale in the lens plane over which the plasma density decreases by a factor of $e$. Thus, the maximum source size, in physical units, for which the image magnification can reach $\langle \mu_{\rm tot} \rangle$ is (using Eqns. \ref{eq:delta_umax}, \ref{eta-AB}),
\begin{eqnarray}
\label{eq:size_from_mag}
& \delta x_{\rm mag}\lesssim {2a\over |\eta|\langle\mu_{\rm tot}\rangle^2}
\approx \\ &
\begin{cases}
4.8\times 10^9\,{\rm cm}\,{a_{13}^3\nu_9^2\over d_{\rm SL,19}{\rm DM}_0} \left({\langle\mu_{\rm tot}\rangle\over 10}\right)^{-2}, & {\rm caustic~A},\\[4pt] 4.5\times 10^{10}\,{\rm cm}\, {a_{13}\over \sqrt{\mathcal{L}_{\alpha}/5}}\left({\langle\mu_{\rm tot}\rangle\over 10}\right)^{-2},& {\rm caustic~B}.
\end{cases} \nonumber
\end{eqnarray}
The lens parameter constrained by an observation of a source with total magnification $\mu_{\rm tot}$ is shown in Fig.~\ref{fig:maxsize}. An even stronger limit on the source size can be placed when lensing-induced spectral fringes are observed, see \S \ref{sec:images_interact}.

For a point source, the range in source position over which the pair of near-caustic images has total magnification larger than 
$\langle\mu_{\rm tot}\rangle$ is
$\Delta u_s(>\langle\mu_{\rm tot}\rangle)
\simeq {2\over |\eta|\langle\mu_{\rm tot}\rangle^2}$.
If source positions are distributed approximately uniformly over a scale $\sim 2u_{\rm s,*}$, the corresponding probability is
\begin{equation}
\label{eq:lensingProbability}
P(>\langle\mu_{\rm tot}\rangle) \simeq {\Delta u_s\over 2u_{s,*}}={1\over |\eta|u_{s,*}\langle\mu_{\rm tot}\rangle^2}.
\end{equation}
Substituting the strong-lens caustic locations and slopes (Eqns.~\ref{eq:uIx}, \ref{eq:muGausslens}, and \ref{caustic2}) gives
\begin{eqnarray}
& P(>\!\langle \mu_{\rm tot}\rangle)\!\approx
\begin{cases}
\frac{e}{2\alpha^2 \langle \mu_{\rm tot}\rangle^2},
& {\rm caustic~A},\\[4pt]
\frac{1}{2\mathcal{L}_{\alpha} \langle \mu_{\rm tot}\rangle^2},
& {\rm caustic~B}.
\end{cases} \nonumber
\end{eqnarray}
High magnification is therefore dominated by caustic B (by a factor of $\sim \alpha^2/e\mathcal{L}_{\alpha}$, i.e. $\sim 10$ at $\alpha=10$ and $\sim 60$ at $\alpha=30$) and has the generic fold probability $P(>\langle\mu_{\rm tot}\rangle)\propto \langle\mu_{\rm tot}\rangle^{-2}$.
Although this produces a power-law high-energy tail, its relative normalization is small, $\sim(2\mathcal{L}_\alpha)^{-1}$ (this applies so long as successive bursts sample the relevant source-plane interval independently, or if transverse motion sweeps the sightline across it). Well monitored repeater energy distributions have at their high end, $dN/dE\propto E^{-\gamma}$ with $\gamma<3$ \citep{Li2021Natur,Kirsten2024,Ould-Boukattine2024}. Therefore, below the intrinsic cutoff, the lensed population only slightly renormalizes the intrinsic one and is unobservable. However, above a sharp cutoff $E_c$, only the lensed contribution remains. It dominates for $E>E_c$ and follows the generic fold slope $dN/dE\propto E^{-3}$, independent of $\gamma$ and suppressed by $\sim(2\mathcal{L}_\alpha)^{-1}$ relative to the intrinsic distribution just below $E_c$. A drop in the energy distribution followed by a steepening to this distinct power-law above it, would therefore be a lensing signature - though one requiring a well-sampled distribution near and above $E_c$ to be detectable. Such a distribution has, so far, not been clearly identified in any of the well sampled FRB sources.

\subsection{Relative transverse motion}
If we know that the source has been magnified by a factor $\sim \langle \mu_{\rm tot}\rangle$ for a time $\tau$, then the lateral velocity of the source relative to the lens can be constrained as $v_{\perp} < |\delta x_{sx}|/\tau$.

Moreover, the perpendicular velocity of the lens will result in a frequency-dependent change in the source's dimensionless position. A constant velocity leads to $u_{s}(t)=u_{s,0}+v_{\perp}t/a$ with some initial $u_{s,0}$. A caustic crossing happens when $u_{s}(t_{\rm *})=u_{\rm s,*}(\nu)$. It then follows that for two frequencies separated by $\Delta \nu$, the time difference between crossings is $\Delta t_{\rm *}=\frac{a}{v_{\perp}}\lvert \frac{du_{\rm 1s,*}}{d\nu}\rvert \Delta \nu$. Plugging in $u_{\rm s,*}(\alpha)$ and $\alpha\propto \nu^{-2}$, we get
\begin{eqnarray}
\label{eq:delt*}
    &  \Delta t_{\rm *} \!\approx\! \left\{\hspace{-2pt}  \begin{array}{ll}\sqrt{\frac{2}{e}} \frac{a\alpha}{v_{\perp}}\frac{\Delta \nu}{\nu}, & {\rm caustic~A} \\
    \frac{a}{v_{\perp}\sqrt{\mathcal{L}_{\alpha}}}\frac{\Delta \nu}{\nu}, & {\rm caustic~B}
    \end{array}\right. \\ & \approx \left\{\hspace{-2pt}  \begin{array}{ll}2\times 10^7\frac{d_{\rm SL,19}\mbox{DM}_0}{a_{13}\nu_9^2 v_{\perp,7}}\frac{\Delta \nu}{\nu}\mbox{ s}, & {\rm caustic~A} \\
    4.5\times 10^5\frac{a_{13}}{v_{\perp,7}}\bigg[\frac{\mathcal{L}_{\alpha}}{5}\bigg]^{-1/2}\frac{\Delta \nu}{\nu}\mbox{ s}, & {\rm  caustic~B}
\end{array}\right. \nonumber
\end{eqnarray}
For order-unity frequency separations, the predicted recurrence delay is weeks to months. A candidate event or lensing-induced rate enhancement in a repeater should therefore migrate through neighboring bands with the chromatic drift of Eq. \ref{eq:delt*}, providing a direct test of the lens interpretation. The drift is much larger near caustic A than near caustic B.

\subsection{Near-critical images}
\label{sec:images_interact}

\subsubsection{Time delays of burst copies and the wave-optics magnification limit}
The phase difference between images is determined by
\begin{equation}
\label{eq:Phi}
\Phi(u)={\pi a^2\over R_F^2}\left[(u_I-u_s)^2-\alpha e^{-u_I^2} \right].
\end{equation}
However, the arrival time of a burst image is the group delay,
$t_g=\partial\Phi/\partial\omega$, evaluated at the stationary point. Since 
$\partial\Phi/\partial u_I=0$ at an image,
\begin{equation}
 \label{eq:group_delay}
t_g(u_I)={a^2\over 2cd_{\rm eff}} \left[(u_I-u_s)^2+\alpha e^{-u_I^2}\right] .
\end{equation}
For a host-galaxy lens, $d_{\rm eff}\simeq d_{\rm SL}$. In Eq. \ref{eq:group_delay}, the sign of the plasma term is opposite to that in the phase, reflecting the distinction between plasma phase delay and group delay.

Using the approximate image locations from \S \ref{sec:gauss_preliminaries}, we first estimate the central (inner) image away from the immediate vicinity of caustic A. For $\alpha\gg1$, $u_I\sim u_s/(1+\alpha)$, and Eq.\ref{eq:group_delay} gives $t_{g,\rm central}\sim \frac{a^2}{2cd_{\rm eff}} \left(u_s^2+\alpha\right)$. When $u_s^2\ll\alpha$ (over much of the three-image region including near caustic B), the plasma term dominates, giving
\begin{equation}
\Delta t_{\rm central}\sim \frac{a^2\alpha}{2cd_{\rm eff}}=\alpha T_G=T_p^{\rm abs}.
\end{equation}
This is dominated by the DM delay of the full lens column density, because the central image passes through the peak of the Gaussian. Near caustic A, by contrast, $u_s\sim\alpha/\sqrt{2e}$, so the geometrical contribution is not negligible; the corresponding third-image delay is treated explicitly below in Eq. \ref{eq:3rd}.

As shown in \S \ref{sec:gauss_preliminaries}, away from the caustic, the central image is strongly demagnified and therefore may be observationally missed. A more typical situation will be to observe two copies of the burst with the delay corresponding to the two non-central images
\begin{equation}
\Delta t_{\rm bright}\sim {a^2\over 2cd_{\rm eff}}\times {\cal O}(u_s^2)=T_G\times {\cal O}(u_s^2),
\end{equation}
with the coefficient depending on the source location within the caustic interval.
Importantly, this is independent of $\alpha$.

Consider now the special case of a source located close to a caustic.  Near a caustic, we take, as before, $u_{I}=u_{I*}+\epsilon$ (with $\epsilon$ positive or negative). The phase difference between the two merging images is cubic in their image-plane separation. Expanding the phase around $u_{I*}$, where both the first and second derivatives vanish, gives
$\Delta\Phi  \equiv \Phi(u_+)-\Phi(u_-)
\simeq {4\over 3}{\pi a^2\over R_F^2} \eta\epsilon^3$. Using $|\epsilon|=\frac{2}{|\eta| \langle \mu_{\rm tot}\rangle}$, the corresponding
phase-delay scale is
\begin{equation}
    {|\Delta\Phi|\over \omega}
    \simeq
    {16a^2\over
    3c d_{\rm eff}\eta^2\langle\mu_{\rm tot}\rangle^3}=\frac{32T_G}{3\eta^2\langle\mu_{\rm tot}\rangle^3}.
    \label{eq:phase_delay_scale}
\end{equation}
This is the phase accumulated between the two image contributions at a fixed frequency, but it should not be identified with the temporal separation of the two burst copies.

The geometrical-optics expression (Eq. \ref{eq:phase_delay_scale}) cannot be extrapolated to arbitrarily large magnification. The two stationary points are well-defined as separate images only when their phase difference satisfies $|\Delta\Phi|\gg1$. Defining the wave-optics transition by $|\Delta\Phi|\sim1$ gives
\begin{equation}
\label{eq:muw0}
    \mu_{\rm w}\equiv  \left({32\pi a^2\over3R_F^2\eta^2}\right)^{1\over 3}\!=\!\left({32\lambda r_eN_0\over3\alpha\eta^2} \right)^{1\over 3}\!=\!\left({32\Phi_0\over3\alpha\eta^2} \right)^{1\over 3}\!=\!\left({32\Phi_G\over3\eta^2} \right)^{1\over 3}.
\end{equation}
For $\langle\mu_{\rm tot}\rangle\gtrsim\mu_{\rm w}$, the stationary points are not separately phase resolved and diffraction washes out the geometrical divergence. $\mu_{\rm w}$ thus gives an approximate scale for the maximum magnification (finite source size and instrumental averaging can reduce the observed peak further).

For the two caustics of a strong Gaussian lens,
\begin{eqnarray}
\label{eq:muw}
   & \mu_{\rm w}\!\approx\! \begin{cases} \left({4e\lambda r_eN_0\over3\alpha^3}   \right)^{1\over 3}\!\approx\! 15\,{\rm DM}_0^{1\over 3}\nu_9^{-{1\over 3}}
    \left({30\over\alpha}\right) & {\rm caustic~A},\\[0pt] \left( {8\lambda r_eN_0\over3\alpha \mathcal{L}_{\alpha}} \right)^{1\over 3}\!\approx \!83\,{\rm DM}_0^{1\over 3}\nu_9^{-{1\over 3}} \left[{120\over \alpha\mathcal{L}_{\alpha}}\right]^{1\over 3} & {\rm caustic~B}.  \end{cases}  
\end{eqnarray}
Thus, at fixed lens strength and caustic geometry, $\mu_{\rm w}\propto{\rm DM}_0^{1/3}$. Equivalently, the minimum lens DM is
\begin{eqnarray}
\label{eq:DM0min}
    \mbox{DM}_0\!\gtrsim \!\begin{cases} 0.3 \left(\frac{\alpha}{30}\right)^3\left(\frac{\langle\mu_{\rm tot}\rangle}{10}\right)^3\nu_9\mbox{pc cm}^{-3} & {\rm caustic~A},\\[0pt] 1.7\times 10^{-3}\frac{\alpha \mathcal{L}_{\alpha}}{120}\left(\frac{\langle\mu_{\rm tot}\rangle}{10}\right)^3\nu_9\mbox{pc cm}^{-3} & {\rm caustic~B}.  \end{cases} 
\end{eqnarray}
The DM difference between the bright pair and third image is of order $\mbox{DM}_0$ for both caustics A and B. Therefore, given a variation $\delta \mbox{DM}$ between bursts from a repeater Eq. \ref{eq:DM0min} directly constrains the maximum allowed magnification\footnote{For $\alpha\to \alpha_{\rm cr}$ the caustics merge and the limit becomes apparently less restrictive, $\langle \mu_{\rm tot} \rangle\lesssim 800 \delta \mbox{DM}^{1/2}\nu_9^{-1/2}$. However, this only holds for a very narrow range of frequencies, $\Delta \nu/\nu\lesssim (4 \langle \mu_{\rm tot} \rangle)^{-1}$}. Since the image delays associated with the lens are uncorrelated with any external sources of DM (which may dominate the DM budget for the source), they are statistically unlikely to reduce the DM variation below DM$_0$. Therefore, large amplification is inconsistent with fluctuations in the total DM, smaller than DM$_0$\footnote{This kind of analysis will naturally be less constraining with regards to repeater sources with highly variable DMs.}.

The observable arrival-time separation of the two near-caustic images is instead the group-delay difference,
\begin{equation}
    \Delta t_g    \equiv t_g(u_+)-t_g(u_-)  =    {\partial \Delta\Phi\over \partial \omega} \approx  -{4a^2\over cd_{\rm eff}}  \alpha u_{I*} e^{-u_{I*}^2}\epsilon.
\end{equation}
where in the last part we expanded Eq. \ref{eq:group_delay} around $u_{I*}$ to leading order in $\epsilon$.
Taking the absolute value, relating $\epsilon$ to $\langle \mu_{\rm tot}\rangle$ and using the critical-curve condition, $\alpha e^{-u_{I*}^2}=1/(2u_{I*}^2-1)$,
we obtain
\begin{eqnarray}
\label{eq:near_caustic_group_delay}
  &  |\Delta t_g|
    \simeq {8a^2\over cd_{\rm eff}}
    {\alpha u_{I*}e^{-u_{I*}^2}
    \over |\eta|\langle\mu_{\rm tot}\rangle} \simeq 
    {8a^2\over cd_{\rm eff}}
    {u_{I*}\over
    |2u_{I*}^2-1|\,|\eta|\,
    \langle\mu_{\rm tot}\rangle}\\
    & \approx \begin{cases}
{2a^2\over cd_{\rm eff}\langle\mu_{\rm tot}\rangle}=\frac{4T_G}{\langle\mu_{\rm tot}\rangle}
& {\rm caustic~A},\\[0pt]
{2a^2\over
    cd_{\rm eff}\mathcal{L}_{\alpha}
    \langle\mu_{\rm tot}\rangle}=\frac{4T_G}{\mathcal{L}_{\alpha}\langle\mu_{\rm tot}\rangle}
& {\rm caustic~B}.
\end{cases} \nonumber
\end{eqnarray}
While the phase-delay scale in Eq. \ref{eq:phase_delay_scale} varies as $\langle\mu_{\rm tot}\rangle^{-3}$, the burst-copy arrival-time separation varies as $\langle\mu_{\rm tot}\rangle^{-1}$.
This difference arises because the group delay is sensitive to the frequency dependence of the plasma phase.
For observations near caustic A, the third (unmagnified) image is the outer one, corresponding to a negative delay and a strong frequency dependence ($\propto \alpha^2 \propto \nu^{-4}$). For caustic B, it is the inner image which is the third (demagnified) one, and it has a positive dispersive delay ($\propto \alpha \propto \nu^{-2}$) relative to the bright near-caustic pair:
\begin{eqnarray}
\label{eq:3rd}
    \mu_{3}\!\approx \!\begin{cases}1 & {\rm caustic~A},\\[0pt]{1\over \alpha} & {\rm caustic~B}. \end{cases} \, t_{g,3}\!\approx \!\begin{cases}-\frac{\alpha^2 a^2}{4e c d_{\rm eff}}\!=\!-\frac{\alpha T_{p}^{\rm abs}}{2e} & {\rm caustic~A},\\[0pt]\frac{\alpha a^2}{2 c d_{\rm eff}}\!=\!T_{p}^{\rm abs} & {\rm caustic~B}. \end{cases}
\end{eqnarray}

\subsubsection{Image interference and spectral modulation}
The locus of the two merging images in the time--frequency plane has a simple form near a frequency-domain caustic crossing. 
Consider a frequency $\nu_*$ that corresponds to a caustic crossing $u_{I}=u_{I*}$ of a static lens. The lens strength at this frequency is $\alpha_*=\alpha(\nu_*)$.
Expanding the lens equation about this point gives
\begin{equation}
    {\nu-\nu_*\over\nu_*}    \simeq     {\eta_*\over4Q_*u_{I*}}\epsilon^2, \qquad    Q_*\equiv\alpha_*e^{-u_{I*}^2}={1\over2u_{I*}^2-1},
\end{equation}
where $\eta_*=\eta(u_{I*},\nu_*)$. The group delay then satisfies $t_g-t_{g,*} \simeq -4T_G Q_*u_{I*}\epsilon$ with $t_{g,*}=t_g(u_{I*},\nu_*)$. Eliminating $\epsilon$,
\begin{equation}
\label{eq:parabola}
    {\nu-\nu_*\over\nu_*} \simeq {\eta_*\over64T_G^2Q_*^3u_{I*}^3} (t_g-t_{g,*})^2 .
\end{equation}
Thus the two image tracks form a parabola near the frequency-domain caustic. Since $\eta_A<0$ and $\eta_B>0$, the parabola opens toward lower frequencies near caustic A and toward higher frequencies near caustic B. Measuring this signature could indicate lensing and constrain a unique combination of lens parameters.

\begin{figure*}
\centering
\includegraphics[width=0.48\textwidth]{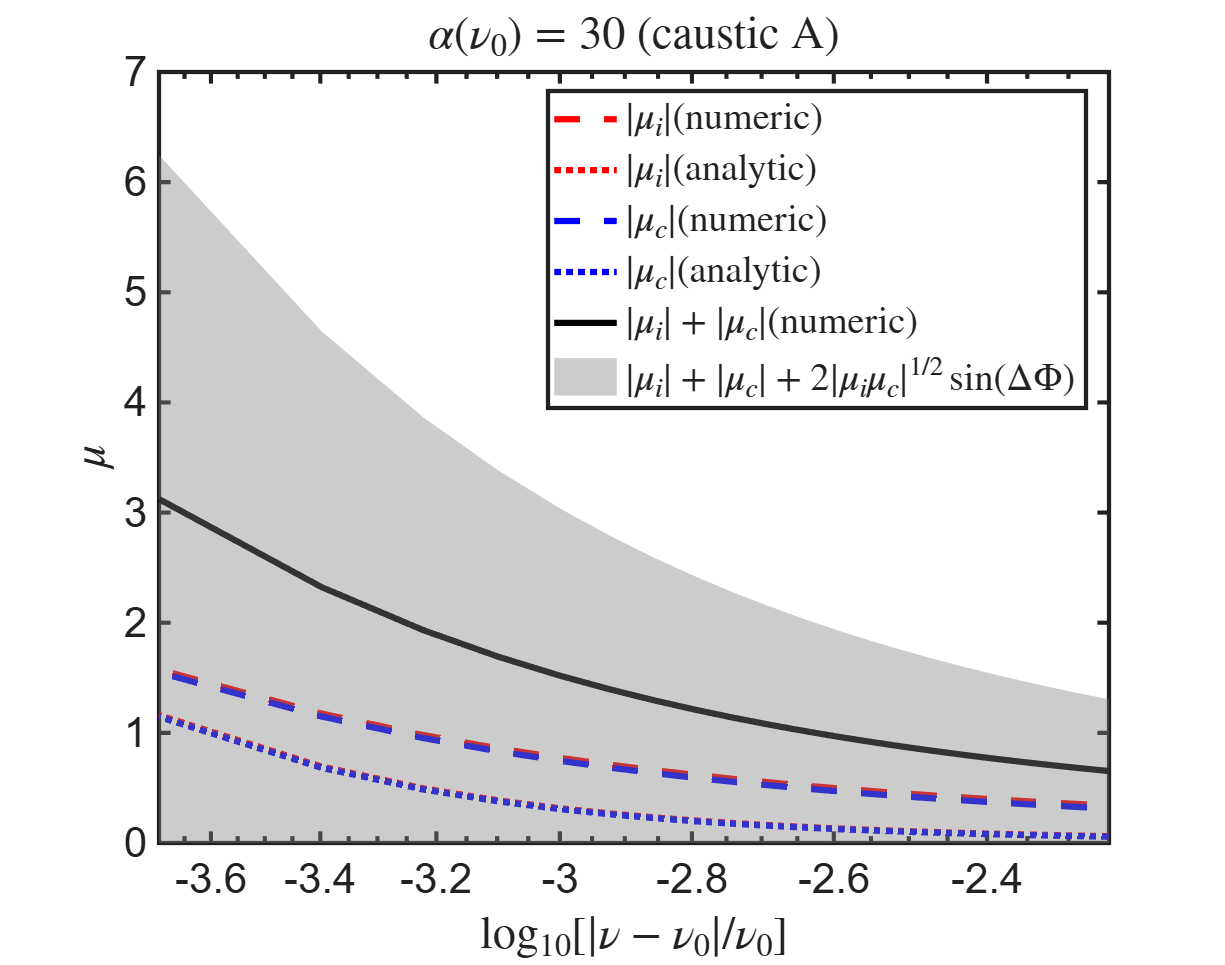}
\includegraphics[width=0.48\textwidth]{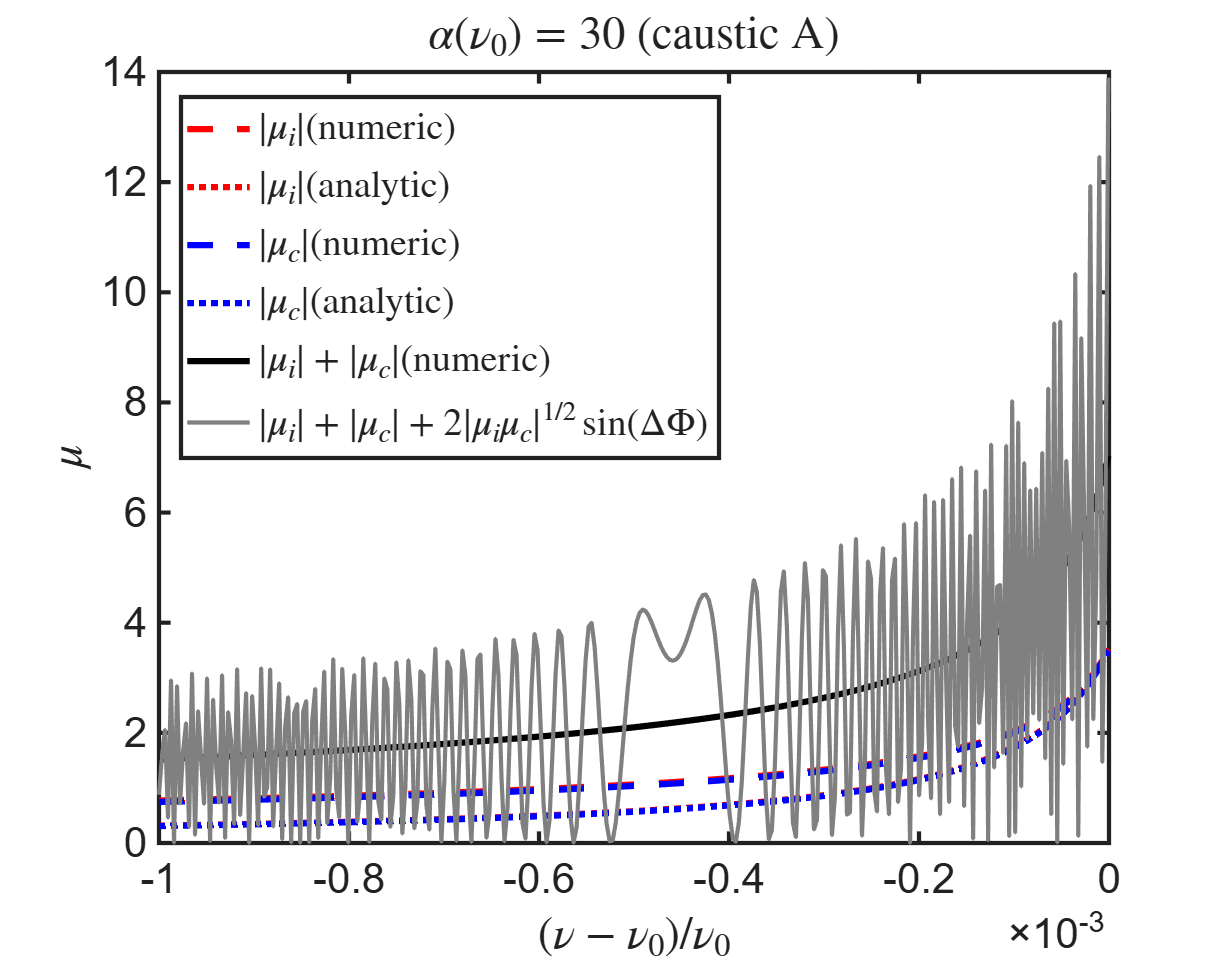}\\
\includegraphics[width=0.48\textwidth]{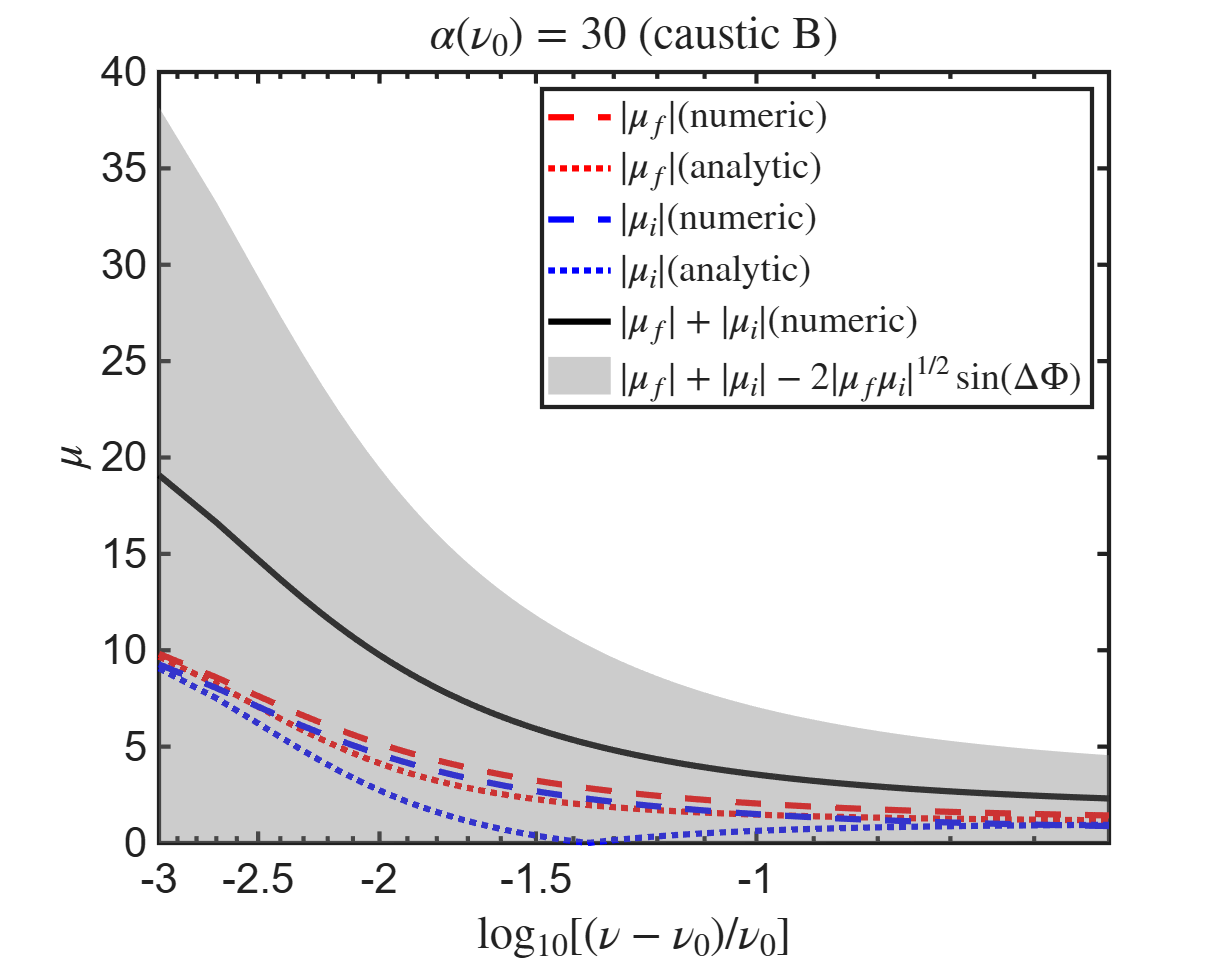}
\includegraphics[width=0.48\textwidth]{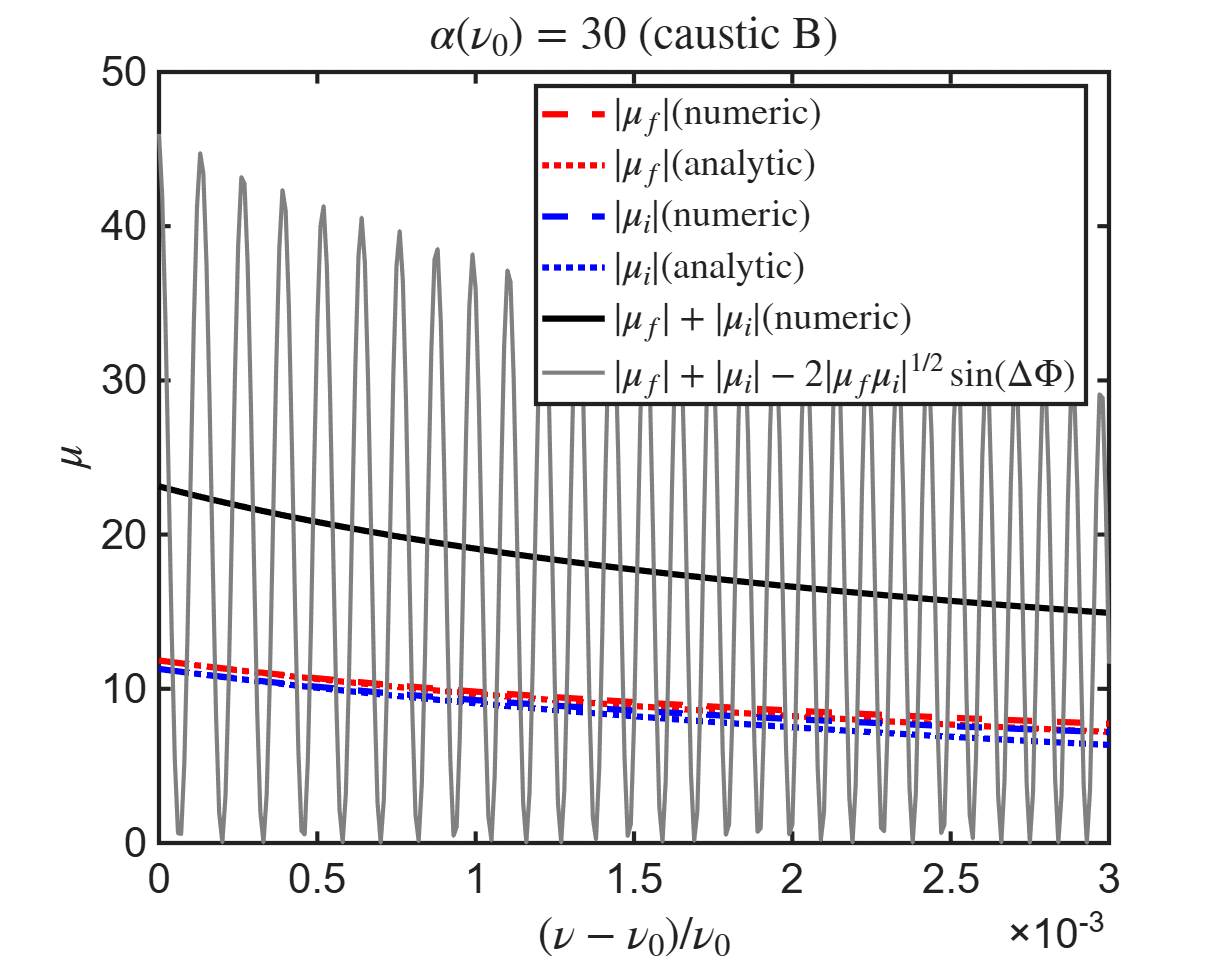}
\caption{The magnification spectrum near caustics A (top) and B (bottom). We assume here $a=10^{13}\mbox{ cm}, d_{\rm SL}=1.24\times 10^{19}\mbox{ cm}\ll d_{\rm LO}, \nu_0=1\mbox{ GHz}, \mbox{DM}_{\rm 0}=1\mbox{pc cm}^{-3}$ as well as $u_1=u_{0,A}-0.001$ for caustic A and $u_1=u_{0,B}+0.001$ for caustic B. Two nearby images contribute large magnification. We show both the full numerical results (dashed lines) and the analytic approximation given by Eq. \ref{eq:Delmuofnufin} (dotted lines). The total spectrum is either the incoherent sum (black), corresponding to $|g^{(1)}(\Delta t_g)|\ll 1$, or the coherent summation of electric fields (grey band), corresponding to $|g^{(1)}(\Delta t_g)|=1$. The latter follows Eq. \ref{eq:mu_partial_coh} and involves a term oscillating with the phase difference between the images (Eq. \ref{eq:phase_delay_scale}).} The envelope of the magnification decreases on a frequency scale $\Delta \nu_{\rm env}$ (Eq. \ref{eq:lensing_bandwidth_general}). The fringe spacing is much smaller, $\Delta \nu_{\rm fr} \sim \Delta t_g^{-1}\ll \nu$ (Eq. \ref{eq:Delta_nu_fringe}) and as such is unresolved in the left panels - it is shown directly in the right panels.  
\label{fig:spectrum}
\end{figure*}

Eq. \ref{eq:delta_umax} relates $\langle \mu_{\rm tot}\rangle$ to $|u_{s}-u_{s,*}|$. The caustic position, $u_{s,*}$ depends on $\alpha$ and hence on $\nu$.
The characteristic fractional bandwidth over which the magnification changes by an order-unity factor is therefore
\begin{eqnarray}
\label{eq:lensing_bandwidth_general}
  &  {\Delta\nu_{\rm env}\over \nu}\!\sim \! \frac{|u_{s}-u_{s,*}|}{\frac{du_{s,*}}{d\log \nu}}\! \sim \!\left|\frac{d u_{\rm s,*}}{d\log\nu}\right|^{-1} \! {2\over |\eta|\langle\mu_{\rm tot}\rangle^2} \! \approx\! 
  \begin{cases} {e\over 2\alpha^2\langle\mu_{\rm tot}\rangle^2} & {\rm caustic~A},\\[0pt]{1\over \langle\mu_{\rm tot}\rangle^2} & {\rm caustic~B}. \nonumber
\end{cases} \\  & \approx\! 
  \begin{cases} 2\times 10^{-5}\nu_9^4 d_{\rm SL,19}^{-2}\mbox{DM}_0^{-2}a_{13}^4\left(\frac{10}{\langle\mu_{\rm tot}\rangle}\right)^{2} & {\rm caustic~A},\\[0pt]0.01\left(\frac{10}{\langle\mu_{\rm tot}\rangle}\right)^{2} & {\rm caustic~B}. 
\end{cases}
\end{eqnarray}
up to ${\cal O}(1)$ logarithmic factors.
This is the instantaneous width of the smooth magnification envelope, assuming a locally smooth intrinsic spectrum. The analytic spectral expansion is given in \S \ref{sec:spectrum} and compared with numerical results in Fig. \ref{fig:spectrum}.

For a point source, the observational appearance of the two near-caustic images depends on the group-delay separation $|\Delta t_g|$.
We denote by $t_{\rm coh}$ the intrinsic coherence time of the emitted FRB electric field, independent of propagation and instrumental filtering. The interference actually observed between the two images is governed by their mutual coherence after propagation and spectral filtering; its relation to $t_{\rm coh}$ is discussed in \S \ref{visibility}. Three useful observational regimes follow.

\begin{figure}[ht]
\centering
\includegraphics[width=0.43\textwidth]{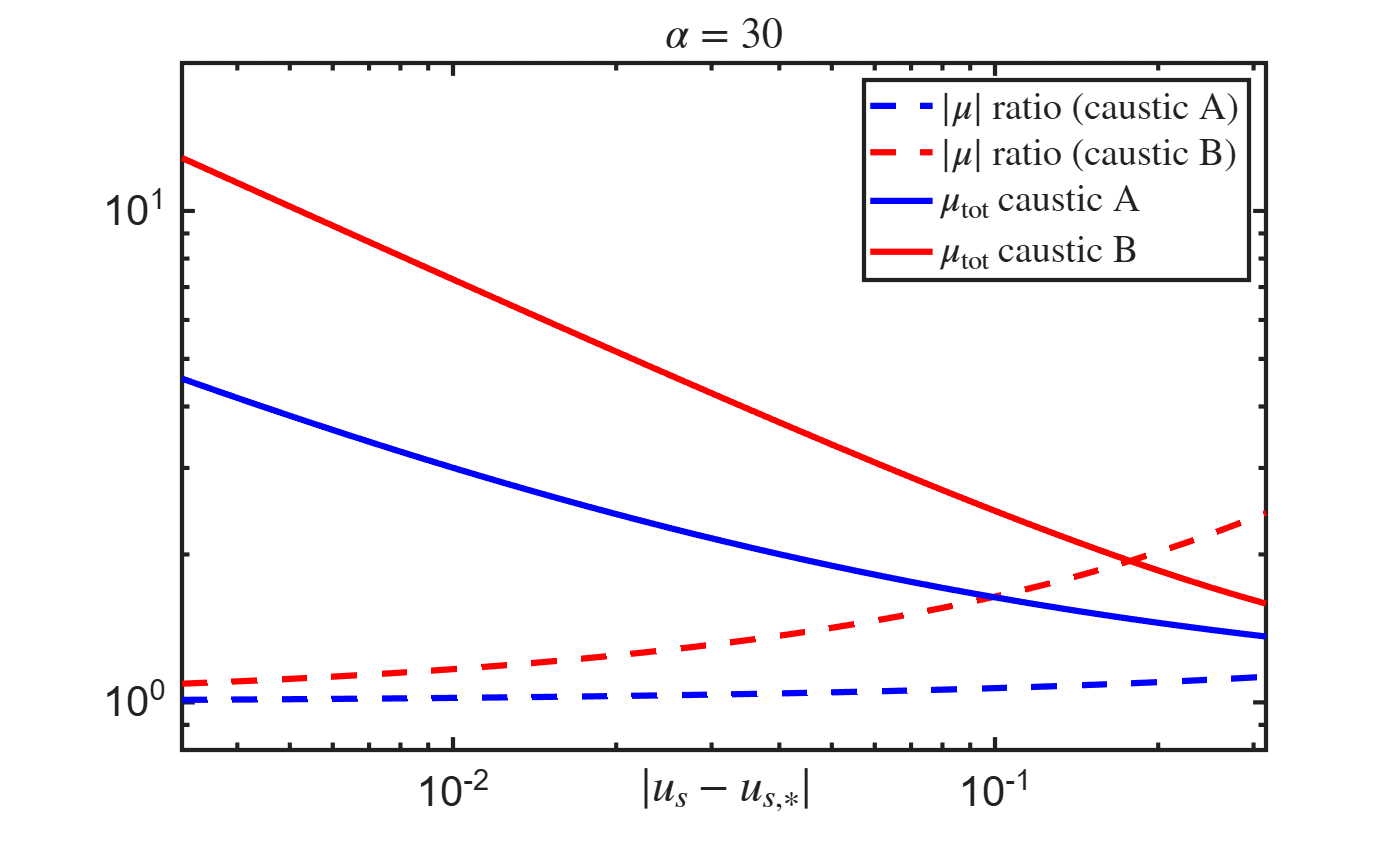}
\includegraphics[width=0.43\textwidth]{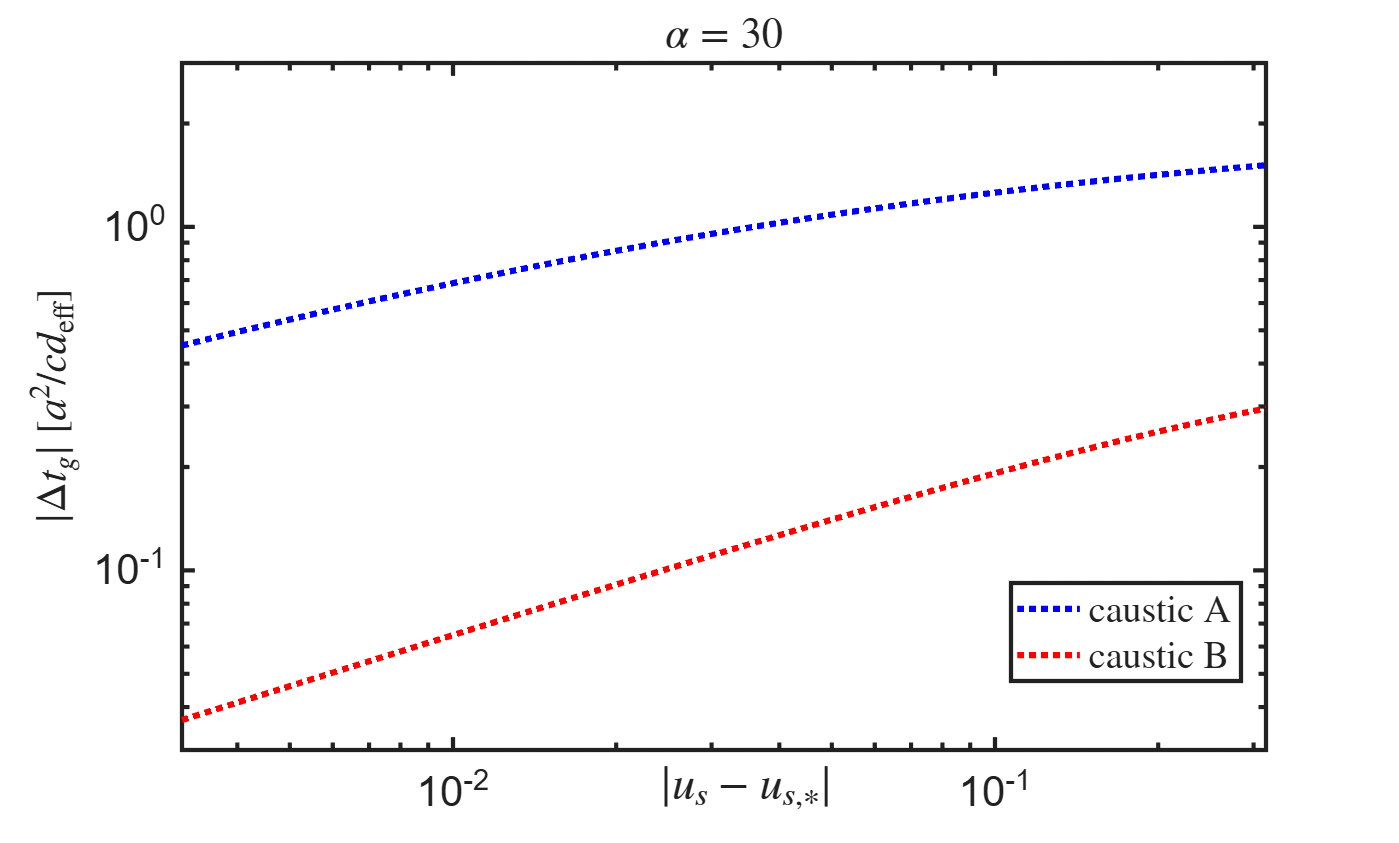}
\caption{Properties of the two merging images near a caustic.
The upper panel shows the magnification ratio and the total incoherent magnification of the two images. The lower panel shows the group-delay separation $|\Delta t_g|$ between the two images, in units of $a^2/(cd_{\rm eff})$. This delay sets the temporal separation of resolved burst copies and the spectral fringe spacing in the coherent regime. } 
\label{fig:twoimages}
\end{figure}

\begin{enumerate}
    \item If $|\Delta t_g|\gtrsim t_{\rm FRB}$, the two magnified images appear as temporally resolved burst copies. Their arrival-time separation, $|\Delta t_g|$, and flux ratio are directly observable (Fig. \ref{fig:twoimages}). Since the image envelopes do not substantially overlap, spectral interference is not visible in the time-integrated burst even if the fields retain mutual coherence.

    \item If the image envelopes overlap but the two image fields have negligible mutual coherence at the spectral resolution of the observation\footnote{I.e., $|\Delta t_g|\lesssim t_{\rm FRB}$ but $|g^{(1)}(\Delta t_g)|\ll1$, see \S \ref{visibility}}, their intensities add approximately incoherently.
    \begin{equation}
    \mu_{\rm inc}=|\mu_+|+|\mu_-|\simeq \langle\mu_{\rm tot}\rangle .
    \label{eq:mu_inc}
    \end{equation} 
    This case is difficult to distinguish from a single magnified image.
    However, the two summed images generally pass through slightly different parts of the plasma lens and sample different electron columns with slightly different dispersive delays. Such a burst cannot, in general, be perfectly de-dispersed with a single DM, and this may provide a diagnostic of unresolved plasma-lensed images.

    For the Gaussian lens, the lens contribution to the DM of an image at position $u_I$ is ${\rm DM}(u_I)={\rm DM}_0 e^{-u_I^2}$. The difference between the near-critical images is
    \begin{eqnarray}
    \label{eq:DMcol}
    &  |\Delta{\rm DM}_{\rm col}|\! \simeq \! 4u_{I*}|\epsilon|\,{\rm DM}_0 e^{-u_{I*}^2} \! \simeq \! {8u_*{\rm DM}_0e^{-u_*^2} \over  |\eta|\langle\mu_{\rm tot}\rangle}  \\
      & \approx \begin{cases} 8.3\times10^{-3}\,{\rm pc\,cm^{-3}}\,
    {a_{13}^2\nu_9^2\over d_{\rm SL,19}} {10\over \langle\mu_{\rm tot}\rangle} & {\rm caustic~A},\\[0pt] 1.7\times10^{-3}\,{\rm pc\,cm^{-3}}\,  {a_{13}^2\nu_9^2\over d_{\rm SL,19}}\frac{5}{\mathcal{L}_{\alpha}}{10\over \langle\mu_{\rm tot}\rangle} & {\rm caustic~B}.
    \end{cases} \nonumber
    \end{eqnarray}
    This column-density difference is not identical to the differential DM inferred from the total arrival-time slope, because the chromatic geometric delay also differs between the two near-caustic images. To leading order, the geometric and plasma contributions to the group-delay difference are comparable, giving an effective differential DM $|\Delta{\rm DM}_{\rm fit}|\simeq 2|\Delta{\rm DM}_{\rm col}|$\footnote{The equality of the geometric and dispersive contributions follows from the stationary-phase condition together with $\Phi_p\propto \omega^{-1}$, and is independent of the lens profile and caustic identity. That being said, the exact coefficient relating $|\Delta{\rm DM}_{\rm fit}|$ to $|\Delta{\rm DM}_{\rm col}|$ depends on how the effective DM is fitted over a finite band.}. Here $\Delta{\rm DM}_{\rm fit}$ is the difference between DMs that would be obtained by fitting each image with a $\nu^{-2}$ arrival time delay. Thus an unresolved, incoherent pair of plasma-lensed images leaves a residual frequency-dependent temporal structure after de-dispersion with any single DM that provides a consistency check independent of the lens profile: two burst copies must satisfy $\Delta t_g=8.3\Delta \mbox{DM}_{\rm col}\nu_9^{-2}{\rm ms}$.

    \item If the image fields retain appreciable mutual coherence\footnote{I.e. $|\Delta t_g|\lesssim t_{\rm FRB}$ and $|g^{(1)}(\Delta t_g)|$ is appreciable}, they interfere. The observed magnification is
    \begin{eqnarray}
    \label{eq:mu_partial_coh}
    & \mu_{\rm obs}(\nu) =|\mu_+|+|\mu_-|+\\ &2|\mu_+\mu_-|^{1/2}\Re\left\{g^{(1)}[\Delta t_g(\nu)] \exp\left[i\left(\Delta\Phi(\nu)+\Delta\phi_{\rm M}\right)\right] \right\}, \nonumber
    \end{eqnarray}
    where $\Delta\phi_{\rm M}$ is the relative stationary-phase, or Maslov, phase ($\Delta\phi_{\rm M}=-\pi/2$ for caustic A and $+\pi/2$ for caustic B). The fringe contrast is affected by temporal coherence, unequal image magnifications, finite source size and spectral resolution, polarization mismatch, and scattering (Appendix \ref{visibility}). If these factors vary slowly across one oscillation, the fringe spacing is set by the frequency derivative of the relative phase, $\Delta\nu_{\rm fr}\simeq|\Delta t_g|^{-1}$, leading to
    \begin{eqnarray}
    \label{eq:Delta_nu_fringe}
      &  \frac{\Delta\nu_{\rm fr}}{\nu} \approx (\nu|\Delta t_g|)^{-1} \\ & \approx \!\begin{cases}
    1.5\times 10^{-5}d_{\rm SL,19}a_{13}^{-2}\nu_9^{-1}\frac{\langle\mu_{\rm tot}\rangle}{10} & {\rm caustic~A},\\[0pt] 7.5\times 10^{-5}d_{\rm SL,19}a_{13}^{-2}\nu_9^{-1}\frac{\mathcal{L}_{\alpha}}{5}\frac{\langle\mu_{\rm tot}\rangle}{10}
    & {\rm caustic~B}.
    \end{cases} \nonumber
    \end{eqnarray}
    The coherence function controls the fringe contrast, whereas $\Delta t_g$ controls the fringe spacing. The small fringe spacing makes them very challenging to detect without baseband/voltage data. Even when such resolution is achievable, a weak or absent fringe pattern does not directly constrain the expected fringe spacing; it may instead reflect $|g^{(1)}(\Delta t_g)|\ll1$ or one of the additional averaging effects in Eq. \ref{eq:visibility}.
    Note that $\Delta\nu_{\rm fr}$ is distinct from $\Delta\nu_{\rm env}$ discussed above.
    The former is set by the group-delay separation of the two images, $\Delta\nu_{\rm fr}\simeq |\Delta t_g|^{-1}$ and describes the oscillatory fringes, whereas the latter is set by the rate at which the caustic moves in the source plane as the observing frequency changes and describes the smooth envelope over which the lens significantly magnifies the burst. Depending on the lens parameters, the envelope may contain many fringes, one fringe, or less than a full fringe.  We define the number of fringes under the envelope as $N_{\rm fr}=\Delta \nu_{\rm env}/\Delta \nu_{\rm fr}$.

    Detection of spectral fringes with appreciable contrast requires both temporal mutual coherence and limited averaging over the source. Even when $|g^{(1)}(\Delta t_g)|$ is large, different source elements sample different relative phases between the two images. For the near-caustic pair,
    \begin{equation}
    \left|{d\Delta\Phi\over du_s}\right|\simeq{8\pi a^2\over R_F^2|\eta|\langle\mu_{\rm tot}\rangle}.
    \end{equation}
    The fringe phase therefore changes by less than order unity across the source only if $\delta u_s \lesssim {R_F^2|\eta|\langle\mu_{\rm tot}\rangle \over8\pi a^2}$. For a host-galaxy lens this corresponds to 
    \begin{eqnarray}
    \label{eq:source_size_fringe}
   & \delta x_{\rm fr} \!\lesssim \!{R_F^2|\eta|\langle\mu_{\rm tot}\rangle \over8\pi a}\!\approx \\&
   \!\begin{cases}
    4.9\times 10^{8}\mbox{ cm}\,d_{\rm SL,19}^2\mbox{DM}_0 a_{13}^{-3}\nu_9^{-3}\frac{\langle\mu_{\rm tot}\rangle}{10} & {\rm caustic~A},\\[0pt] 5.4\times 10^{7}\mbox{ cm}\,\frac{d_{\rm SL,19}}{a_{13}\nu_9}\sqrt{\frac{\mathcal{L}_{\alpha}}{5}}\frac{\langle\mu_{\rm tot}\rangle}{10}
    & {\rm caustic~B}.
    \end{cases}. \nonumber
    \end{eqnarray}
    Eq. \ref{eq:source_size_fringe} is an order-unity estimate. The more precise relation between source size and fringe visibility for a specified brightness profile is given in \S \ref{visibility}.
    
    In the geometrical-optics regime this source-size limit is more restrictive than the finite-source condition required to obtain the observed magnification, and their ratio is inversely proportional to $N_{\rm fr}$: $\delta x_{\rm fr}/\delta x_{\rm mag}\approx \frac{2}{3}(\langle \mu_{\rm tot}\rangle/\mu_{\rm w})^3\sim \frac{1}{10N_{\rm fr}}$ (with the numerical coefficient corresponding to $\alpha=30$ and caustic B). A heavily fringed spectrum requires $\mu_{\rm tot}\ll\mu_{\rm w}$, whereas the highest magnified events should show a smooth envelope with $\lesssim 1$ fringe.
\end{enumerate}

\begin{figure}
\centering
\includegraphics[width=0.47\textwidth]{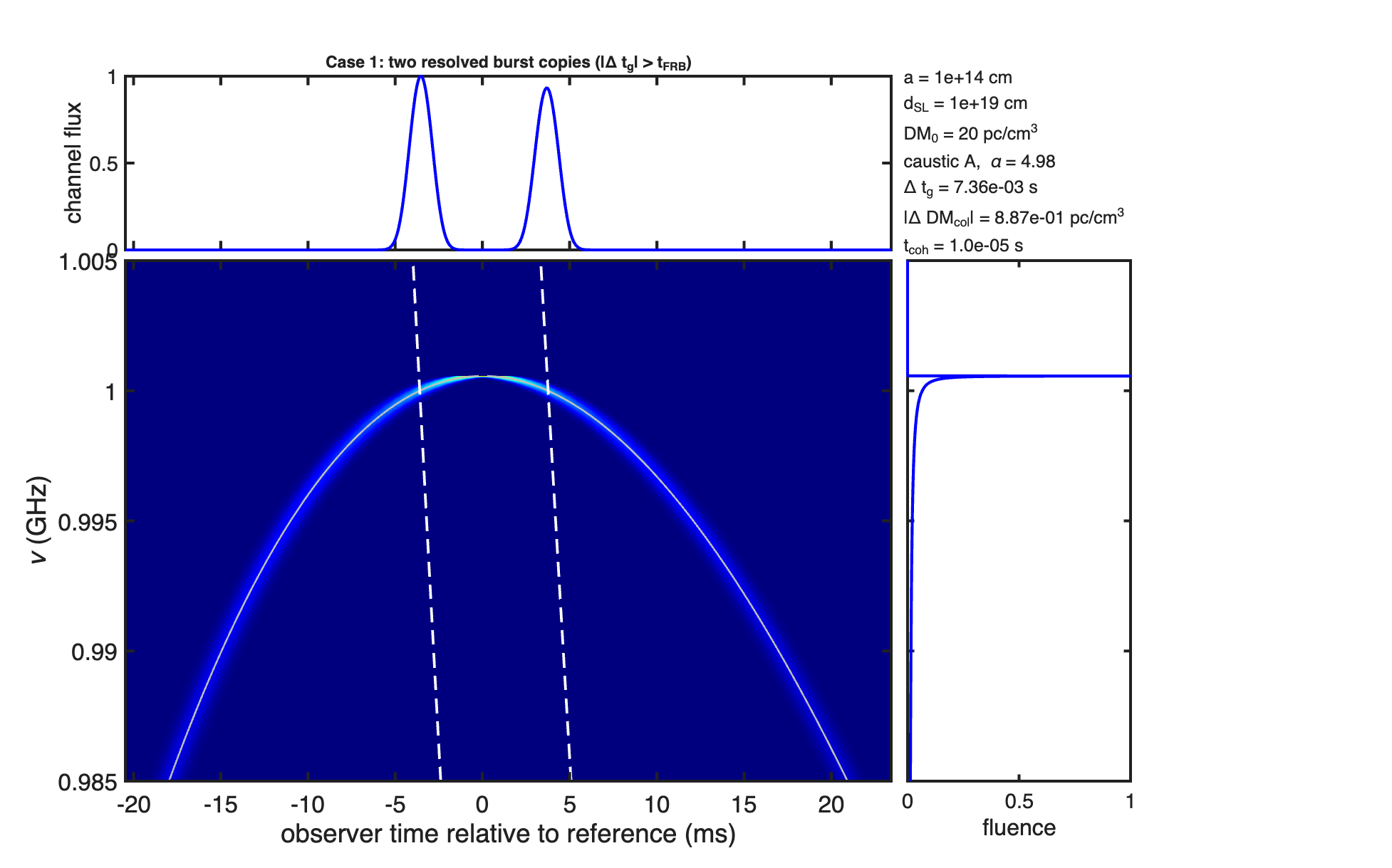}\\
\includegraphics[width=0.47\textwidth]{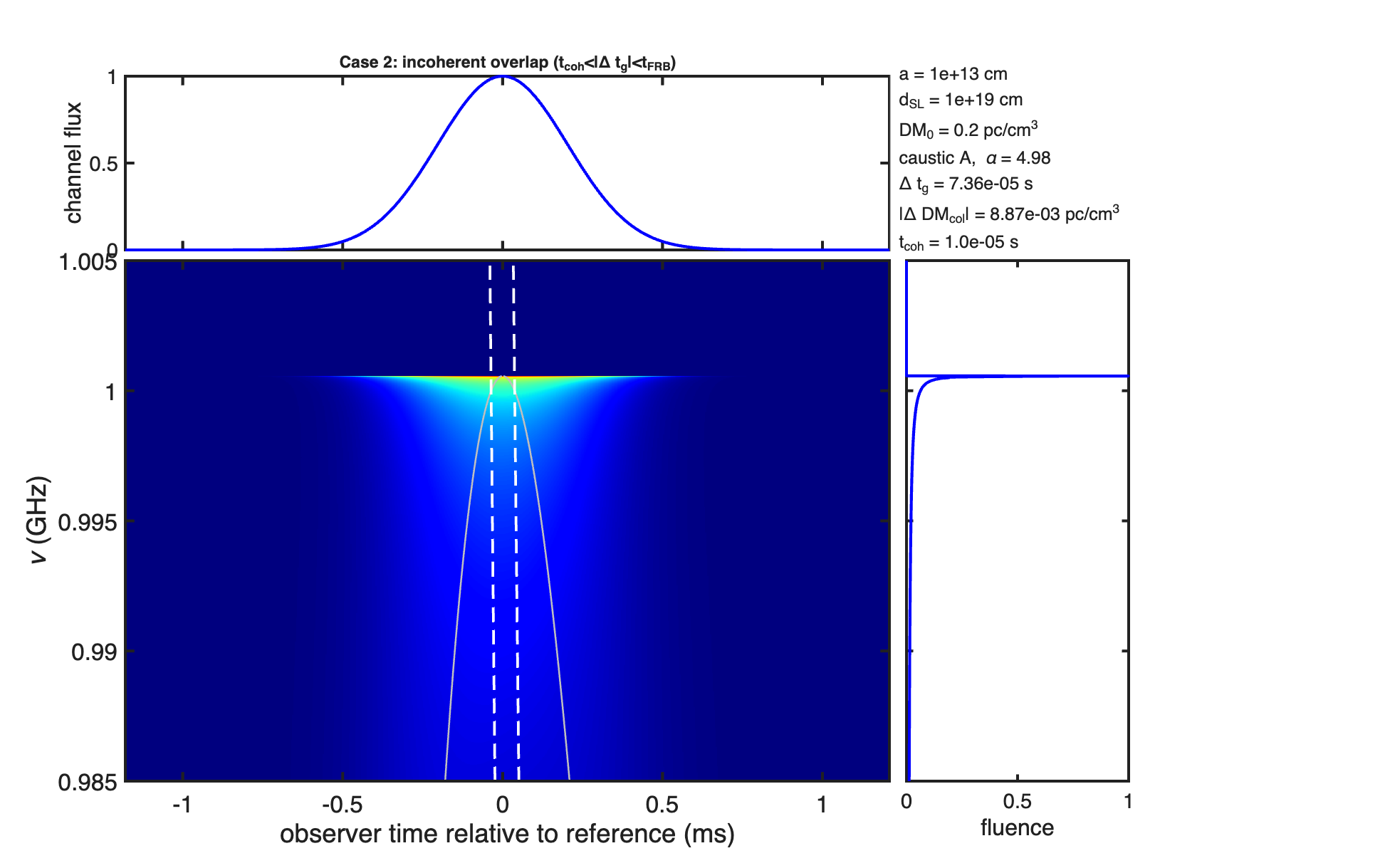}\\
\includegraphics[width=0.47\textwidth]{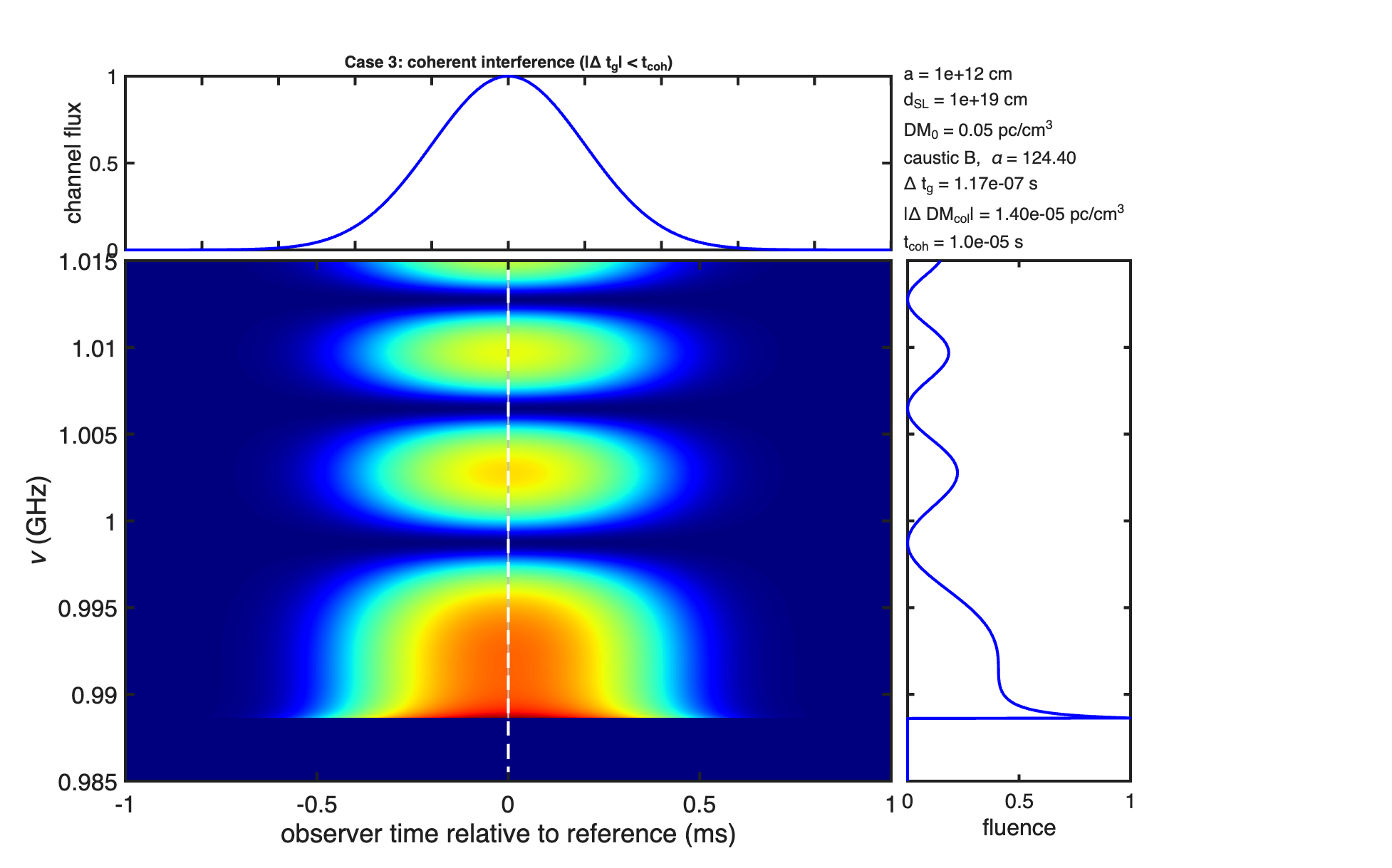}
\caption{Dynamic spectra illustrating the three observational regimes of a near plasma-lens caustic image pair (\S \ref{sec:images_interact}). No dedispersion has been applied; the maps show received intensity vs. observer time and frequency, after subtraction of a frequency-independent reference time. Upper sub-panels show the temporal profile in a narrow channel around $\nu=1$\,GHz, while right-hand sub-panels show time-integrated spectral fluence. Top: $|\Delta t_g|>t_{\rm FRB}$, so the two images appear as temporally resolved burst copies. Middle: the image envelopes overlap and $|g^{(1)}(\Delta t_g)|\ll1$, so the images add approximately incoherently and produce residual chromatic structure that cannot be described exactly by a single cold-plasma dispersion law. Bottom: $|g^{(1)}(\Delta t_g)|$ is appreciable, so the image fields interfere and produce spectral fringes. The example shown assumes $|g^{(1)}|=1$; partial coherence would reduce the fringe contrast without changing the fringe spacing. Solid curves trace the exact group-delay tracks of the near-caustic images. Dashed curves show the corresponding $\nu^{-2}$ cold-plasma tracks based on the electron column sampled at the reference frequency, illustrating the additional chromatic contribution from the frequency-dependent image positions. Near the frequency at which the source crosses the caustic, the two image tracks merge along the local fold-caustic relation $(\nu-\nu_*)/\nu_*\propto (t_g-t_{g,*})^2$; the parabola opens toward lower (higher) frequencies for caustic A (B). Only the two images associated with the selected caustic are shown.}
\label{fig:waterfall}
\end{figure}

Magnifications and time delays vs caustic distance are plotted in Fig. \ref{fig:twoimages}. Waterfall plots demonstrating cases 1-3 described above are presented in Figure \ref{fig:waterfall}.

\subsection{Magnetized lenses}
\label{sec:magnetized}
If the plasma lens is magnetized, the different images may sample different rotation measures. Plasma lensing may then produce substantial frequency-dependent depolarization, apparent circular polarization, or large apparent PA changes \citep{LiLensing2026}. For this to occur, a necessary condition is that the difference in the electric field rotation angle is large, $\Delta \chi_{\rm col}=\lambda^2\Delta \mbox{RM}_{\rm col}>1$ (e.g. \citealt{BKN2022}).

For a Gaussian lens, the differential DM between the two near-caustic images is given by Eq. \ref{eq:DMcol}. If the line-of-sight magnetic field has approximately constant magnitude and sign between the two ray paths,
\begin{equation}
\label{eq:deltaRM}
 |\Delta{\rm RM}_{\rm col}| \simeq 0.812\,B_{\parallel,\mu{\rm G}} |\Delta{\rm DM}_{\rm col}| \ {\rm rad\,m^{-2}}.
\end{equation}

For the two near caustic images the condition $\Delta \chi_{\rm col}>1$ then reduces to
\begin{eqnarray}
\label{eq:Bmin0}
    B_{\rm min}\! \approx \!\begin{cases}
    1.6\,{\rm mG}\, d_{\rm SL,19}a_{13}^{-2} \bigg(\frac{\langle \mu_{\rm tot}\rangle}{10}\bigg) & {\rm caustic~A},\\[0pt] 8.3\,{\rm mG}\, d_{\rm SL,19}a_{13}^{-2}\frac{\mathcal{L}_{\alpha}}{5}\bigg(\frac{\langle \mu_{\rm tot}\rangle}{10}\bigg)  & {\rm caustic~B}.
    \end{cases}
\end{eqnarray} 
Because the near-caustic ray separation decreases as $\langle\mu_{\rm tot}\rangle^{-1}$, the required field grows linearly with magnification and is larger near caustic B. For the canonical dimensional parameters $d_{{\rm eff},19}=a_{13}=\nu_9={\rm DM}_0=1$, corresponding to $\alpha\simeq24$, and for $\langle\mu_{\rm tot}\rangle=10$, order-unity differential rotation requires fields of a few mG.

\subsection{Relativistic corrections}
Because the phase is Lorentz invariant, the plasma phase may be computed in the comoving frame of the lens. There are two main cases of interest. First, consider the case in which a lens is a localized clump comoving with a relativistic wind moving radially outward from the source. The radio wave and the plasma move in the same direction, and $\omega'=\gamma(1-\beta)\omega\simeq{\omega\over2\gamma}$.
Therefore $\Phi_p=-{\xi N'_e\over \omega'}=-{2\gamma\,\xi N'_e\over\omega}$, where $N'_e$ is the proper electron column through the lens. The effective observed DM of the moving lens (i.e. the column reproducing the observed phase through the static cold-plasma relation) is therefore ${\rm DM}_{\rm eff}\simeq 2\gamma\,{\rm DM}_0$, where DM$_0$ is defined in the plasma comoving frame. Equivalently, for a fixed proper column, the lensing strength is enhanced to $\alpha_{\rm eff}\simeq 2\gamma\,\alpha_0$. 
Second, consider the DM contributed by the extended relativistic wind in which such a lens is embedded. In this case the outward propagating photon is moving only marginally faster than the wind (their relative velocity being $c(1-\beta)$), and only overtakes a column $N_{\rm cross}\approx (2\gamma^2)^{-1}\int n d\ell$ (with $n$ and $\ell$ measured in the lab-frame). For example, for $n\propto r^{-2}$ this gives $\sim n_0 R_0/(2\gamma^{2})$ - only the region close to the base contributes because the density decreases faster than the photon can overtake it (the result remains qualitatively unchanged so long as $n(r)$ declines faster than $r^{-1}$). Combined with the $2\gamma$ enhancement of the phase, we get DM$_{\rm eff}=\gamma^{-1}\int n d\ell$.
Importantly, when the lens contribution to the observed DM is measured directly, the relativistic correction is already contained in that measurement, and $\alpha$ follows Eq.  \ref{eq:alpha} with no further factor of $\gamma$. The correction matters only when converting an observed lensing strength into a physical density or mass-loss rate for the lensing structure.

A large normalized wave amplitude, $a_0\gg1$, modifies the plasma response by increasing the effective electron inertia. To order of magnitude, the plasma frequency is reduced as $\omega_p^2\rightarrow {\omega_p^2\over \gamma_q}$ with $\gamma_q\sim a_0$, so the plasma phase and the effective DM are suppressed by $\gamma_q^{-1}$. Combining this with the bulk relativistic correction gives ${\rm DM}_{\rm eff} \simeq {2\gamma\over \gamma_q}\,{\rm DM}_0$. As above, for a fixed observed DM contribution, however, the lensing strength $\alpha$ inferred from that observed DM is unchanged; the nonlinear correction only changes the proper column required to produce the observed phase.

\section{Observables and their implications}
\label{sec:Observe}
\subsection{Comparison with observations and search strategies}
As described in \S \ref{sec:intro}, lensing cannot be securely established from one observable. Intrinsic spectral structure, sub-burst drift, scintillation, and chance similarities can mimic narrow spectra, chromatic tracks, or repeated burst morphologies. A credible identification should fit several correlated observables with one lens model; these are summarized in Table~\ref{tab:lensing_observables}. For instance, the lens parameters fit by \cite{Uttarkar2026} for FRB 20240114A, based on the rate enhancement, fix $a^2/d_{\rm eff}$ and imply a maximum image delay $\Delta t_{g\rm, max}\lesssim 10$\,ms, more than three orders of magnitude below the minute to hour separations of their reported ``carbon copies".

From an observational perspective, a useful search sequence is the following. Candidate events may be identified via multiple burst copies, unusually narrow magnification envelopes, or paired time--frequency tracks. The image tracks should then be fit jointly, rather than independently dedispersed, and tested for the frequency-dependent delay and differential DM predicted by the same lens solution. If sufficiently fine spectral data are available, one should search for fringes with $\Delta\nu_{\rm fr}\simeq|\Delta t_g|^{-1}$ inside the broader magnification envelope. A search for the third image and for correlated polarization structure provides additional tests. Finally, for repeaters, continued multi-band monitoring can test the predicted chromatic migration of the event. A non-detection of any one of these features is constraining only when the available time resolution, spectral resolution, bandwidth, and polarization sensitivity are sufficient to detect the corresponding prediction.

The most diagnostic signature would be a pair of image tracks that merge at a frequency-domain caustic and obey the parabolic relation given by Eq. \ref{eq:parabola} (with the curvature direction different between caustics A and B), together with the associated magnification enhancement. Resolved burst copies should exhibit frequency-dependent delays, flux ratios, and apparent DMs that evolve consistently with the same lens model. Detection of the third image would provide an additional strong test: near caustic A an approximately unmagnified image precedes the bright pair, whereas near caustic B a demagnified inner image follows it. Crucially, near caustics, all observables are powers of a single variable, the distance from the caustic $\epsilon$, or equivalently, $\langle \mu_{\rm tot}\rangle$. This is illustrated in Table~\ref{tab:lensing_observables}. 
A lensing sequence predicts $\langle \mu_{\rm tot}\rangle \Delta t_g=const, \langle \mu_{\rm tot}\rangle^2\Delta \nu_{\rm env}=const, \langle \mu_{\rm tot}\rangle\Delta {\rm DM}=const$, independent of the detailed fold profile. If changes in observed burst energy ($E_{\rm obs}$) are dominated by lensing rather than intrinsic burst-to-burst variations, or within copies of a given lensed burst, the corresponding relations may be tested using $E_{\rm obs}$ as a magnification proxy.
For caustic B, Eq. \ref{eq:lensing_bandwidth_general} can be inverted to give a simple relation, independent of lens parameters, $\langle\mu_{\rm tot}\rangle=\sqrt{\nu/\Delta \nu_{\rm env}}$. In other words, a 10\% fractional spectral width corresponds to amplification by a factor of $\sim 3$. Combining this with the magnification probability $P(>\mu)\propto \mu^{-2}$ (Eq. \ref{eq:lensingProbability}) gives $P(<\Delta \nu/\nu)\propto \Delta \nu/\nu$. If the observed energy variations are lensing-dominated, this also predicts an anti-correlation $\Delta \nu/\nu \propto E^{-2}$ which can be tested against data to check if narrow spectral width is due to lensing (for caustic A, $\Delta \nu/\nu\sim 10^{-5}$ and the associated probability is tiny, so it cannot account for prevalent narrow spectra).

In the coherent regime, a particularly powerful signature is the simultaneous presence of a smooth magnification envelope of width $\Delta\nu_{\rm env}$ and fine fringes with $\Delta\nu_{\rm fr}\simeq|\Delta t_g|^{-1}$. These two scales constrain different combinations of the lens parameters and therefore overdetermine the model (moreover $\langle \mu_{\rm tot}\rangle/\Delta \nu_{\rm fr}=const$ independently of the lens model). Since the expected fringes may have widths of only tens of kHz at GHz frequencies, searches should preferentially use stored voltage data, coherent dedispersion, and channelization much finer than standard survey products. Full-Stokes voltage data also permit a search for polarization changes synchronized with the lensing fringes or image overlap.

Repeating FRBs are especially useful to test lensing. Relative transverse motion causes the caustic crossing to drift in time with frequency according to Eq.~\ref{eq:delt*}. A lensing event detected in one band should therefore recur in neighboring bands with a predictable time offset and should be accompanied by correlated changes in fluence, burst-detection rate, bandwidth, and eventually a demagnification trough. Such a joint prediction is substantially more restrictive than identifying isolated narrow-band bursts or apparently similar burst pairs. Additionally, $\alpha\propto \nu^{-2}$ means there is a critical frequency $\nu_{\rm cr}\equiv \nu(\alpha=\alpha_{\rm cr})$,
\begin{equation}
    \nu_{\rm cr}=3.3d_{\rm SL,19}^{1/2} \mbox{DM}_0^{1/2}a_{13}^{-1}\mbox{ GHz}
\end{equation}
Near-caustic lensing is only observable at $\nu<\nu_{\rm cr}$. 

The lens-strength condition can be phrased in an environment-independent form. Writing the column as $N_0=nL$ for a structure of density $n$ and LOS thickness $L$, and requiring $\alpha>\alpha_{\rm cr}$ 
we get a limit on the lens ratio
\begin{equation}
\label{eq:aspect}
    \frac{a}{d_{\rm eff}} \lesssim 6\times10^{-6}\,\nu_9^{-1}\left(\frac{n}{{\rm cm^{-3}}}\right)^{1/2}\left(\frac{L}{d_{\rm eff}}\right)^{1/2}.
\end{equation}
For a localized lens with LOS thickness $L\lesssim d_{\rm eff}$, strong lensing requires structures that are transversely thin relative to their distance from the source unless the density is extremely high. The dependence on $L/d_{\rm eff}$ quantifies the advantage of elongation: for a quasi-isotropic clump ($L\sim a$) the condition tightens to $a/d_{\rm eff}\lesssim 3.6\times 10^{-11}\nu_9^{-2}(n/\mbox{ cm}^{-3})$. Sheets, filaments, and shock interfaces are  favored both geometrically and, as shown in \S\ref{sec:lensing_from_turbulence}, because they avoid producing many competing images. We use the same variables to write the polarization requirement of \S\ref{sec:magnetized}. The condition
$\Delta\chi_{\rm col}>1$ reduces to
\begin{equation}
B_{\rm min} \approx 0.7\ {\rm mG}\, \left(\frac{\alpha}{10}\right) \left(\frac{\langle\mu_{\rm tot}\rangle}{10}\right) \nu_9^{2}\left(\frac{{\rm DM}_0}{{\rm pc\,cm^{-3}}}\right)^{-1}
\label{eq:Bmin}
\end{equation}
near caustic A, and a factor of a few larger near caustic B. 

Table~\ref{tab:environments} lists typical parameters for candidate environments. High density can compensate for a small effective distance: young supernova-remnant filaments and the interface between the ejecta and a magnetar-wind nebula satisfy Eq.~\ref{eq:aspect} for $a \lesssim 10^{13}$--$10^{14}$~cm and exceed $B_{\rm min}$ by $1-2$ orders of magnitude. A magnetar-wind nebula is, however, pair-dominated, so any appreciable ${\rm RM}$ requires a substantial ion contribution. This is most plausible near the interface with the surrounding remnant, where the wind has swept up baryonic ejecta. Binary wind-collision shells formally satisfy both conditions most easily, and could produce events tied to orbital phase, but only with columns implying ${\rm RM}_0\sim10^{8}\ {\rm rad\,m^{-2}}$, three orders of magnitude above the largest measured for any FRB and large enough to depolarize the burst within realistic channel widths. Such lenses are therefore expected to be rare. At the opposite extreme, extended \hbox{H\,\textsc{ii}} regions and Galactic screens easily satisfy the column requirement but fall short of $B_{\rm min}$ by factors of $\sim2$ and $\sim4\times 10^{3}$ respectively, and are more likely to yield a confused network of caustics (\S\ref{sec:lensing_from_turbulence}). Correlated polarization structure in a lensing candidate thus points to a compact circumburst lens rather than a foreground screen. In addition, Galactic screens only produce lensing with very short group time delays, $\Delta t_{g,\rm max}\lesssim 0.2 \mbox{ ms}<t_{\rm FRB}$, making them unlikely to produce separated burst copies. The condition becomes more stringent for higher magnifications. Since $\langle \mu_{\rm tot}\rangle\lesssim \Delta t_{g, \rm max}/\Delta t_g$ (Eq. \ref{eq:near_caustic_group_delay}), resolved copies require $\langle \mu_{\rm tot}\rangle<\Delta t_{g,\rm max}/t_{\rm FRB}$. As a result, well-separated, highly magnified burst copies strongly favor compact circumburst lenses over Galactic screens (independent of polarization considerations mentioned above).

\begin{deluxetable*}{llll}
\tablecaption{Observable signatures of near-caustic plasma lens, magnification scaling, and the lens or source properties that they constrain. The table can be used as a consistency checklist for candidate plasma-lensing events. \label{tab:lensing_observables}}
\tablewidth{\textwidth}
\tablehead{
\colhead{Observable} &
\colhead{Eq.} & \colhead{$\propto \mu^p$} &
\colhead{Main constraint}
}
\startdata
Pair delay, $\Delta t_g$ & 
\ref{eq:near_caustic_group_delay} & -1 & $a^2/d_{\rm eff}$ given the caustic identity and $\mu$. 
\\
Spectral fringe spacing, $\Delta\nu_{\rm fr}$ &
\ref{eq:Delta_nu_fringe} & 1 &
Equivalent to $\Delta t_g$ because $\Delta\nu_{\rm fr}=\Delta t_g^{-1}$. \\
 & & & Requires mutually coherent image fields, compact source, and fine spectral resolution.
\\
Third-image delay, $\Delta t_{g,3}$ & \ref{eq:3rd} & 0 & $\alpha$ once $a^2/d_{\rm eff}$ is inferred from the near-caustic pair.\\
Third-image magnification, $\mu_3$ & \ref{eq:3rd} & 0 & Additional estimate of $\alpha$ near caustic B and a test of the image identification. \\
Spectral-envelope width, $\Delta \nu_{\rm env}$ & \ref{eq:lensing_bandwidth_general} & -2 & A combination of $\alpha$, $\langle \mu_{\rm tot}\rangle $, and caustic identity. \\
Caustic parabola in $t-\nu$ & \ref{eq:parabola} & 0 & Combination of $a^2/d_{\rm eff}$ and $\alpha$. Sign of the curvature distinguishes caustics A and B. \\
Maximum magnification, $\mu_{\rm w}$ & \ref{eq:muw} & - & Lower limit on the lens column density, ${\rm DM}_0$, at fixed $\alpha$ and geometry. \\
High fold magnification & \ref{eq:size_from_mag} & -2 & Upper limit on projected source size required to attain the observed magnification. \\
Fringe visibility & \ref{eq:visibility_equal_images},\ref{eq:visibility},\ref{eq:source_size_fringe},\ref{eq:visibilisrc} & 1 & Joint upper bound on source size and constraint on mutual coherence at $\Delta t_g$, \\
 & & & given observed spectrum and
instrumental bandpass. \\
Differential DM, $\Delta {\rm DM}_{\rm col}$ & \ref{eq:near_caustic_group_delay},\ref{eq:DMcol} & -1 & $\Delta {\rm DM}$ between images. Directly related to group delay: $\Delta t_g=8.3\Delta{\rm DM}_{\rm col}\nu_9^{-2}\mbox{ ms}$  \\
& &  & for any merging pair. A consistency test of the image identification.\\
Differential RM $\Delta {\rm RM}_{\rm col}$ & \ref{eq:near_caustic_group_delay},\ref{eq:deltaRM} & -1 & $B_{\parallel}\approx 10\Delta {\rm RM}_{\rm col}\frac{1\mbox{ ms}}{\Delta t_g} \nu_9^{-2} \mu{\rm G}$ independent of lens geometry \\
Polarization changes & \ref{eq:Bmin0},\ref{eq:Bmin} &  1 & Magnetic field in the lens. For quasi-uniform field, $\Delta{\rm RM}\simeq 0.812B_{\parallel,\mu{\rm G}}\Delta{\rm DM}_{\rm col}$. \\
Temporal evolution & $\sim a/v_{\perp}$ & 0 & An independent transverse-velocity estimate separates $a$ from $d_{\rm eff}$. \\
$\nu$-dependent caustic-crossing time & \ref{eq:delt*} & 0 &  $a\alpha/v_{\perp}$ (caustic A) or $\sim a/v_{\perp}$ (caustic B).\\
\enddata
\end{deluxetable*}

\begin{deluxetable*}{lcccccccccc}
\tabletypesize{\footnotesize}
\tablecaption{Candidate plasma-lensing environments. We list representative values of the effective lensing distance $d_{\rm eff}$, electron density $n$, relative line-of-sight thickness $L/d_{\rm eff}$, and magnetic field $B_\parallel$. Derived quantities are the lens column ${\rm DM}_0 = nL$, the associated rotation measure ${\rm RM}_0 = 0.812\,B_{\parallel,\mu\rm G}{\rm DM}_0$, the maximum transverse lens scale permitting caustic formation $a_{\rm max}$ (Eq.~\ref{eq:aspect}, evaluated at $\alpha = \alpha_{\rm cr}$), the corresponding maximum group-delay separation, $\Delta t_{g,\rm max}=2a_{\rm max}^2/(d_{\rm eff}c)=4T_G$ (obtained at $\langle\mu_{\rm tot}\rangle\approx 1$), the minimum field $B_{\rm min}$ required for $\Delta \chi_{\rm col}\approx 1$ (Eq.~\ref{eq:Bmin}) and the differential RM between near caustic images (Eq. \ref{eq:deltaRM}). All values assume $\nu = 1$~GHz; $B_{\rm min}$ assumes $\alpha = 10$ and $\langle\mu_{\rm tot}\rangle = 10$ and $\Delta \mbox{RM}_{\rm col}$ assumes $\Delta t_g=0.1$\,ms or equivalently $\Delta \mbox{DM}_{\rm col}=0.012\mbox{pc cm}^{-3}$. The fiducial $\Delta t_g=0.1$\,ms assumes $\langle \mu_{\rm tot}\rangle \approx \Delta t_{g,\rm max}/\Delta t_{g}$ which ranges from $\gtrsim 2$ for Galactic screens to $\gtrsim 10^3$ for the circumburst rows; in the latter case this approaches $\mu_{\rm w}$ and the entry should be read as the $\Delta \mbox{RM}_{\rm col}\propto \Delta t_g$ scaling extrapolated to a common delay rather than a directly attainable configuration.
\label{tab:environments}}
\tablehead{
\colhead{Environment} &
\colhead{$d_{\rm eff}$} &
\colhead{$n$} &
\colhead{$L/d_{\rm eff}$} &
\colhead{$B_\parallel$} &
\colhead{${\rm DM}_0$} &
\colhead{${\rm RM}_0$} &
\colhead{$a_{\rm max}$} &
\colhead{$\Delta t_{g,\rm max}$} &
\colhead{$B_{\rm min}$} &
\colhead{$\Delta \mbox{RM}_{\rm col}$} \\
\colhead{} &
\colhead{(cm)} &
\colhead{(cm$^{-3}$)} &
\colhead{} &
\colhead{($\mu$G)} &
\colhead{(pc\,cm$^{-3}$)} &
\colhead{(rad\,m$^{-2}$)} &
\colhead{(cm)} &
\colhead{(s)} &
\colhead{($\mu$G)} &
\colhead{(rad\,m$^{-2}$)}
}
\startdata
SNR filament (host) & $10^{19}$ & $10^{2}$ & $0.1$ & $10^{2}$ & $30$ & $3\times10^{3}$ & $2\times10^{14}$ & $0.2$ & $20$  & $1$  \\
Ejecta--MWN interface  & $10^{17}$ & $10^{4}$ & $0.1$ & $10^{3}$ & $30$ & $3\times10^{4}$   & $2\times10^{13}$ & $0.2$ & $20$ & $10$ \\
Binary wind-collision shell & $10^{13}$ & $10^{9}$ & $0.1$ & $3\times10^{5}$ & $300$     & $8\times10^{7}$ & $6\times10^{11}$ & $2$ & $2$ & $3\times 10^{3}$\\
\hbox{H\,\textsc{ii}} region (host) & $10^{21}$ & $10$ & $10^{-2}$ & $10$            & $30$ & $3\times10^{2}$  & $2\times10^{15}$ & $0.2$ & $20$  & $0.1$ \\
Galactic ESE sheet & $3\times10^{21}$ & $10^{2}$ & $3\times10^{-7}$ & $5$  & $0.03$    & $0.1$ & $10^{14}$ & $2\times 10^{-4}$ & $2\times10^{4}$ & $5\times 10^{-2}$ \\
\enddata
\tablecomments{Densities and fields are order-of-magnitude estimates and should be
regarded as illustrative; $a_{\rm max}\propto n^{1/2}$ and $B_{\rm min}\propto {\rm DM}_0^{-1}$. 
At $\alpha=\alpha_{\rm cr}$, we get the simple result $\Delta t_g=\Delta t_{g,\rm max}\approx 7.4\mbox{ ms}\mbox{DM}_0\nu_9^{-2}$ (the pair delay does not exceed about twice the dispersive delay, because $\Delta \mbox{DM}_{\rm col}^{\rm max}$ is limited by DM$_0$).
$\Delta \chi_{\rm col}\approx 1$ requires $B_\parallel > B_{\rm min}$. $\Delta \mbox{RM}_{\rm col}$ varies by 5 orders of magnitude between the rows making it a highly discriminating observable. The huge ${\rm RM}_0$ implied by the densest environments may itself cause substantial intra-channel depolarization, which would suppress the observed polarized signal. The binary-wind entry illustrates that satisfying the lens-strength and differential-Faraday conditions does not guarantee observational viability: the implied total RM may strongly depolarize the burst.}
\end{deluxetable*}

\subsection{Comparison to previous source size constraints from plasma lensing}
Our source-size constraints refer to specific observables and differ from earlier notions of plasma-lens resolution. \cite{Cordes2017} obtained a scale $\sim R_F^2/a$ by requiring that source displacement doesn't appreciably change the full Kirchhoff diffraction phase. This is a sufficient point-source condition for preserving fine wave-optical structure, but is not a universal requirement for geometrical magnification by an incoherent extended source.

\cite{Main+2018} treated the coherently contributing lens region as a sharply bounded aperture. Its diffraction width gives
\begin{equation}
 x_{\rm res}\simeq 1.9\left(\frac{\lambda d_{\rm eff}}{\pi}\right)^{1/2} \mu^{-1/2}
\end{equation}
for a linear lens, and a $\mu^{-1/4}$ scaling for a circular lens. By contrast, a smooth fold obeys $\mu\propto|\Delta u_s|^{-1/2}$, so the high-magnification source-plane width scales as $\delta x_{\rm mag}\propto\mu^{-2}$. Preserving two-image fringes imposes the separate differential-phase condition in Eq. \ref{eq:source_size_fringe}.

\section{Extreme lensing events in volume-filling turbulent scattering screens}
\label{sec:lensing_from_turbulence}
We now ask whether a volume-filling turbulent screen can produce a clean event dominated by one plasma lens. 
An identifiable plasma lens may itself contain smaller-scale density fluctuations that scatter the radiation in addition to producing the coherent large-scale deflection. As shown in \S\ref{app:internal_scattering}, separate burst images survive provided that the internal scattering broadening satisfies $\tau_{\rm sc,int}\lesssim|\Delta t_g|$. For the canonical parameters, this requires only $\ell_{\phi,\rm int}/a\gtrsim10^{-6}$, so recognizable burst copies do not require the lens to be smooth on scales comparable to its overall size. Regular high-contrast fringes impose the stronger condition $\tau_{\rm sc,int}\ll|\Delta t_g|$ and require sufficiently correlated scattering along the two image paths.

Consider a Kolmogorov cascade with fluctuations of
\begin{equation}
 \delta n_e(\ell)=n_e(\ell/\ell_{\rm max})^{1/3} 
\end{equation}
where $n_e$ is the mean density of the region and $\ell_{\rm max}$ is the scale at which energy is injected to maintain the turbulence. The excess plasma phase associated with passage of an EM wave through a blob of size $\ell$ is 
\begin{equation}
  \delta \Phi_p(\ell)=r_e\lambda \ell \delta n_e(\ell)=\Lambda \ell \bigg(\frac{\ell}{\ell_{\rm max}}\bigg)^{1/3} 
\end{equation}
where $\Lambda\!\equiv \! r_e \lambda n_e$. The accumulated phase through many eddies of size $\ell$ adds as a random walk and is given by $\Delta \Phi_p(\ell)=\delta \Phi_p(\ell) (L/\ell)^{1/2}$ where $L>\ell_{\rm max}$ is the total size of the turbulent plasma filled region. The diffraction scale, $\ell_{\phi}$, is the eddy scale for which the accumulated phase equals unity $\Delta \Phi_p(\ell)=1$ \citep{BK2020}, leading to
\begin{equation}
\label{eq:ellphi}
    \ell_{\phi}=\frac{\ell_{\rm max}^{2/5}}{\Lambda^{6/5} L^{3/5}}\approx 4\times 10^{9}\nu_9^{6/5}n_e^{-6/5}L_{\rm pc}^{-1/5} \bigg(\frac{\ell_{\rm max}}{L}\bigg)^{2/5}.
\end{equation}
This is valid so long as the diffraction length is larger than the minimum eddy size in the region, $\ell_{\phi}>\ell_{\rm min}$. When this does not occur, the relevant scale becomes $\ell_{\rm min}$ instead. The requirement for strong scintillation is $\max(\ell_{\rm min},\ell_{\phi})<R_F$.

We would like to compare the diffraction and turbulence scales with the required length scale of fluctuations that can lead to significant lensing. As shown in \S \ref{sec:Gausslens}, a single over-dense region of size $\ell$ acts as a plasma lens and the deflection angle from it is given by $\delta \theta(\ell)=(\lambda/2\pi)\partial_x\delta \Phi_p(\ell)\approx (\lambda/\pi) \Lambda (\ell/\ell_{\rm max})^{1/3}$. The corresponding focal length is $|f_{\rm foc}|=\ell/\delta\theta(\ell)$ and the dimensionless convergence is 
\begin{equation}
\label{eq:alphaofell}
    \alpha(\ell)=\frac{d_{\rm eff}}{|f_{\rm foc}|}=\frac{\Lambda \lambda d_{\rm eff}}{\pi \ell_{\rm max}^{1/3} \ell^{2/3}}.
\end{equation}
Although the accumulated phase increases with eddy size, $\alpha(\ell)\propto\ell^{-2/3}$: smaller eddies have weaker columns but larger curvature and are stronger lenses.
We define a characteristic lensing scale, $a^*$, such that $\alpha(a^*)=\alpha_{\rm cr}$, leading to:
\begin{equation}
    a^*\approx \bigg(\frac{d_{\rm eff}\lambda \Lambda}{\pi \alpha_{\rm cr} \ell_{\rm max}^{1/3}}\bigg)^{3/2}\approx 2\times 10^7\nu_9^{-3} d_{\rm eff,kpc}^{3/2} n_e^{3/2}\ell_{\rm max,pc}^{-1/2}\mbox{ cm}
\end{equation}
Strong plasma lensing corresponds to $\alpha>\alpha_{\rm cr}$ and therefore to $\ell<a^*$ (equivalently we may write $\alpha(\ell)=\alpha_{\rm cr} (\ell/a_*)^{-2/3})$. Note that for ISM screens $\ell_{\rm min}\sim 10^7-10^{10}\mbox{ cm}$, and therefore for such screens we may get $a^*<\ell_{\rm min}$, meaning that no single eddy can lead to significant lensing. Since $a^*\propto d_{\rm eff}^{3/2}$ lensing is more prominent in extragalactic turbulent screens.

A simple relation between $a^*$ and $\ell_{\phi}$ is given by
\begin{eqnarray}
    {a^*}^{2/3} \ell_{\phi}^{5/6}=\frac{R_F^{2}}{\pi \alpha_{\rm cr}L^{1/2}},
\end{eqnarray}
independent of the density normalization and hence $\Lambda$.

Since for typical parameters we get $a^*\ll L$, there could be many eddies of size $a^*$ within the turbulent region. This has two important implications. First, the contribution from any single eddy of size $a^*$ is small compared to the accumulated phase by similar eddies along the LOS. This differs significantly from the setup in \S \ref{sec:Gausslens}, and will not lead to a clean lensing event. Second, the fact that there are many eddies of size $a^*$ means that there could be rare alignments of multiple eddies of this size which act as a larger and stronger coherent lens. We turn to explore this possibility next.

\begin{figure*}
\centering
\includegraphics[width=0.45\textwidth]{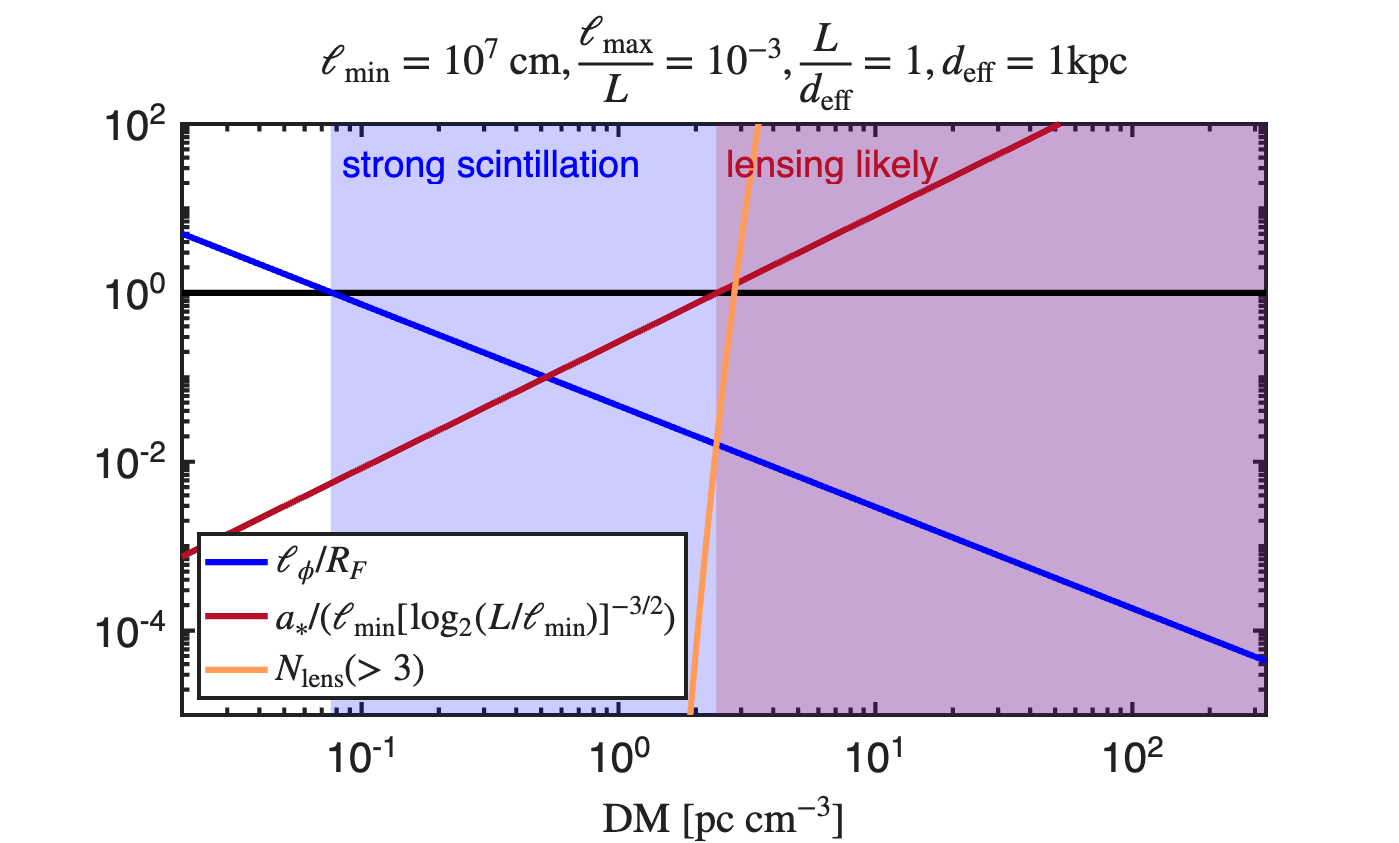}
\includegraphics[width=0.45\textwidth]{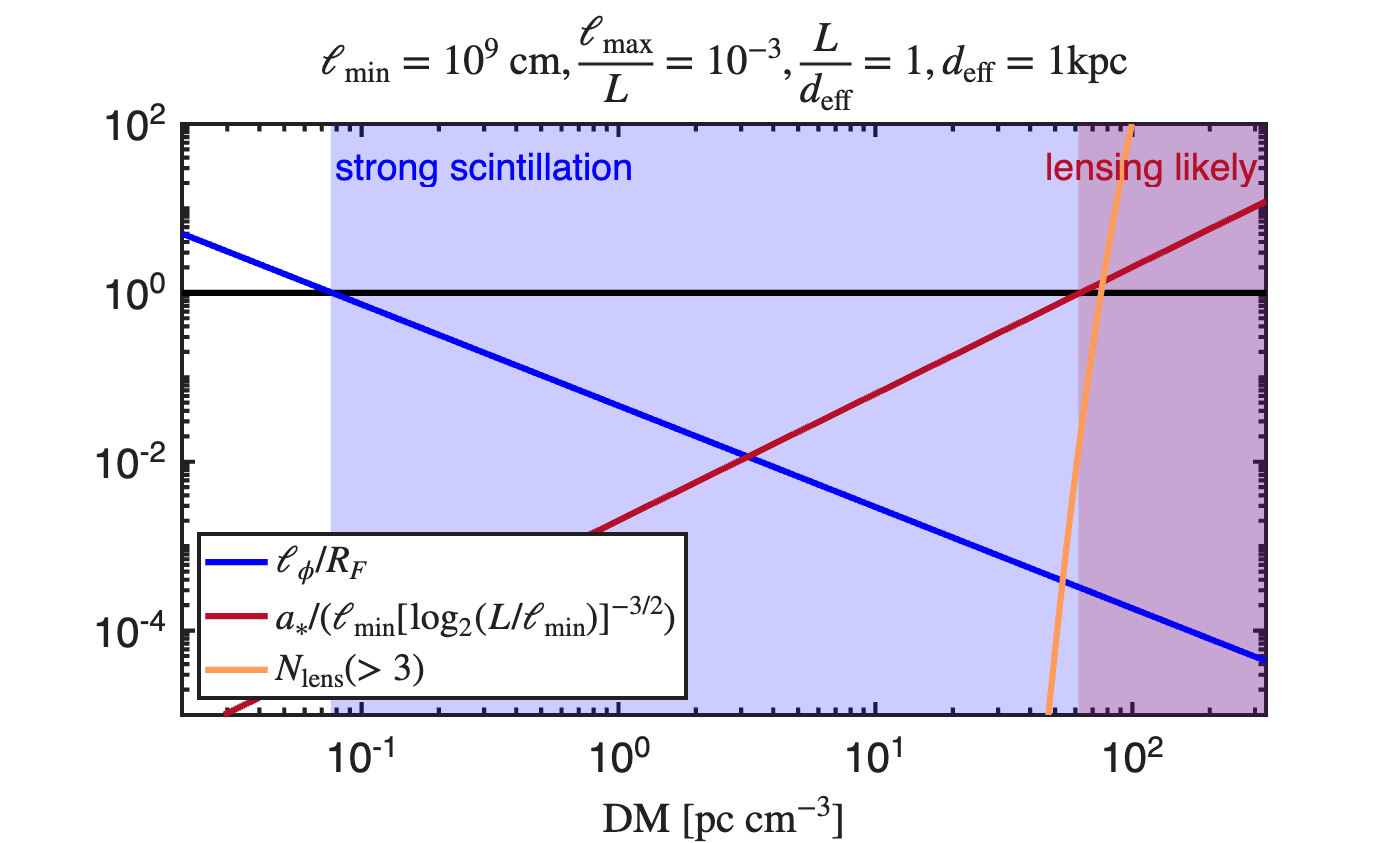}\\
\caption{Parameter space depicting conditions for strong lensing in volume-filling scintillation screens. Results are shown as a function of the DM contribution of the screen. Likely lensing corresponds to $a^*>\ell_{\rm min}[\log_2(L/\ell_{\rm min})]^{-3/2}$ (red shaded region) while strong scintillation requires $\ell_{\phi}<R_{\rm F}$ (blue shaded region). An orange line shows the expected number of lenses along the LOS that contribute to a magnification by a factor $>3$, $N_{\rm lens}(>3)$. Screens that satisfy the conditions for lensing, typically reside in the strong scintillation regime. Since $N_{\rm lens}(>\mu)$ is very sensitive to the DM, there is an extremely narrow region in DM space for which it is neither $\ll 1$ (corresponding to no lensing) nor $\gg 1$ (corresponding to the `confusion' limit).
} 
\label{fig:lensing_scintillation}
\end{figure*}

We consider the case in which $k$ consecutive eddies of size $\ell\gtrsim a^*$ (individually too weak to form caustics) all add their phases coherently, i.e. $\alpha_{\rm eff}(k,\ell)=k\alpha(\ell)$ rather than $\alpha=\sqrt{k}\alpha(\ell)$ as would be expected from random walk. Each eddy can be an over or under density fluctuation, so the probability of each of them being over-density fluctuations is of order $2^{-k}$. The number of opportunities for such a chain to occur is of order $\sim L/\ell$. 
This is an optimistic estimate: In addition to having the same sign, the fluctuations must be sufficiently aligned in transverse position and curvature to act as one coherent lens. The probability $2^{-k}$ therefore neglects geometric penalties and should be regarded as an upper bound on the occurrence of effective chains.
Altogether, the expected number of coherent chains is $N_{\rm ch}(\ell,k)\approx (L/\ell)2^{-k}$. The largest statistically probable chain is the one which occurs on average once along the line of sight. Setting $N_{\rm ch}(\ell,k)=1$ we get the maximum expected $k$ and chain length:
\begin{equation}
k_{\rm ch}(\ell)\sim \log_2(L/\ell)   \quad ; \quad \ell_{\rm ch}\approx \ell k_{\rm ch}(\ell)
\end{equation}
Due to the logarithmic nature of $k_{\rm ch}(\ell)$ it is typically $30-60$ for a large range of screen parameters. Using $k_{\rm ch}(\ell)$, the dimensionless convergence becomes $\alpha_{\rm ch}=\alpha_{\rm cr} k_{\rm ch}(\ell) (a^*/\ell)^{2/3}$. Setting $\alpha_{\rm ch}=\alpha_{\rm cr}$ we get the modified critical lensing scale, $a_{\rm ch}^*\sim {k_{\rm ch}^*}^{3/2}a^*$, where $k_{\rm ch}^*=k_{\rm ch}(a^*)$. So we conclude that coherent chains can occur on scales larger by 2-3 orders of magnitude relative to $a^*$. Accounting for the possibility of such chains, the critical condition for the possibility of strong lensing becomes $\alpha_{\rm ch}(\ell_{\rm min})>\alpha_{\rm cr}$ or $a^*>\ell_{\rm min}[\log_2(L/\ell_{\rm min})]^{-3/2}$. The parameter space for strong lensing by scintillation screens is shown in Fig. \ref{fig:lensing_scintillation}, in comparison to the scintillation properties of the same screens.

The expected number of density fluctuations that lead to strong lensing depends on both the expected number of large chains of eddies resulting in $\alpha_{\rm ch}>\alpha_{\rm cr}$ ($N_{\rm ch}(\ell,k)$) and the geometric probability that the line of sight to such a chain would result in strong lensing, $P_{\rm geo}(\mu,k,\ell) =(2\mathcal{L}_{\alpha_{\rm ch}})^{-1}\mu^{-2}$ (see Eq. \ref{eq:lensingProbability}),
\begin{equation}
\label{eq:Nlensgeneral}
    \langle N_{\rm lens}(>\mu)\rangle \approx \frac{\mu^{-2}}{2\mathcal{L}_{\alpha_{\rm ch}}(k,\ell)}\frac{L}{\ell} 2^{-k}.
\end{equation}
Eq.\ref{eq:Nlensgeneral} is simplified by noting that it is maximized for the lowest values of $\ell,k$ that lead to caustics, i.e. $\ell_{\rm min}$ and $k=\max[1,(\ell_{\rm min}/a^*)^{2/3}]$, ensuring that $\alpha_{\rm ch}\geq \alpha_{\rm cr}$. This leads to two regimes\footnote{In deriving Eq.\ref{eq:Nlens}, $k$ has been treated as a continuous variable. More precisely, when $a^*<\ell_{\min}$ the required chain length is $k_{\rm crit}=\left({\ell_{\min}\over a^*}\right)^{2/3}$, rounded upward to the nearest integer. If $k_{\rm crit}\gtrsim \log_2(L/\ell_{\min})$, even the largest statistically expected coherent chain is unlikely to reach $\alpha_{\rm cr}$. 
Moreover, Eqns \ref{eta-AB}, \ref{eq:lensingProbability} are asymptotic in $\alpha \gg 1$. Near threshold we evaluate $\eta$ numerically from $f(u_{I*})=-1$; the asymptotic expressions overestimate $|\eta|$ by a factor $\lesssim 2$ at $\alpha\approx 2\alpha_{\rm cr}$ and therefore underestimate $P(>\mu)$. Strictly, as $\alpha\to \alpha_{\rm cr}$ the two folds merge into a cusp, with $\mu \propto |\delta u_s|^{-2/3}$ and $P(>\mu)\propto \mu^{-3/2}$. The fold scaling is retained for $\langle \mu_{\rm tot}\rangle \gg 1/2\delta$ with $\delta=\alpha/\alpha_{\rm cr}-1$, so for the thresholds considered here the correction is restricted to $\delta\lesssim 0.2$ and does not affect our conclusions.}:
\begin{eqnarray}
\label{eq:Nlens}
  &  \langle N_{\rm lens}(>\mu)\rangle \approx \frac{\mu^{-2}}{2\mathcal{L}_{\alpha_{\rm ch}}}\frac{L}{\ell_{\rm min}}\left\{ \begin{array}{ll}2^{-(\ell_{\rm min}/a^*)^{2/3}} & ; a^*<\ell_{\rm min} \\ 0.5 & ; a^*>\ell_{\rm min}. \end{array}\right.
\end{eqnarray}

There are three distinct propagation regimes. First, strong scintillation occurs when $\ell_{\rm diff}<R_F$, where $\ell_{\rm diff}\equiv\max(\ell_{\min},\ell_\phi)$. This condition indicates that many random phase patches contribute to the observed field. Second, strong plasma lensing becomes possible when at least one coherent eddy or chain of eddies has $\alpha_{\rm ch}>\alpha_{\rm cr}$. This is a necessary condition for caustic formation, but not by itself sufficient for a clean, isolated lensing event. Third, whether the event is clean or confused is determined by the expected number of caustic-producing structures with magnification larger than a chosen threshold: $\langle N_{\rm lens}(>\mu)\rangle$.
For $\langle N_{\rm lens}(>\mu)\rangle\ll1$, large magnification events are possible but rare. For $\langle N_{\rm lens}(>\mu)\rangle\sim 0.01$--$1$, a detected event is plausibly dominated by a single extreme lens and the analysis of \S\ref{sec:Gausslens} may be applicable. However, as visible in Fig. \ref{fig:lensing_scintillation}, this corresponds to a fine-tuned situation. For $\langle N_{\rm lens}(>\mu)\rangle\gg1$, many such lenses contribute simultaneously; this is the confusion limit, in which the observed signal is better described as strong scintillation or a network of caustics rather than as a single isolated Gaussian lens.

\section{Conclusions}

Plasma lensing can produce many features observed in FRBs, including strong chromatic magnification, narrow spectral structure, multiple images, anomalous arrival-time tracks, interference fringes, and polarization changes. On their own, these signals are insufficient to prove the presence of a plasma lens. Establishing that several observable features arise self-consistently from the same lens configuration is critical for a reliable association. 

For a one-dimensional Gaussian lens, we derived the relations among the principal observables near a fold caustic. The group-delay separation of the two bright images determines either the spacing of resolved burst copies or, when the images retain mutual coherence, the spectral fringe spacing. This delay differs from the phase-delay scale that controls the transition to wave optics and the approximate maximum magnification, $\mu_{\rm w}$. The smooth spectral envelope and the much finer interference fringes therefore probe different combinations of the lens parameters. We also distinguished the finite-source conditions associated with large magnification ($\delta x_{\rm mag}\propto\mu^{-2}$) and with preservation of coherent fringes ($\delta x_{\rm fr}\propto\mu$). 
High-contrast fringes simultaneously constrain the source size and, given the observed spectrum and instrumental bandpass, the temporal coherence of the FRB radiation. In particular, the source must be sufficiently compact to avoid phase averaging across the emitting region. For noise-like emission, however, fine channelization can itself provide coherence over the image delay, so the coherence constraint must be interpreted together with the intrinsic spectral structure and the measurement bandpass.

A robust lensing identification should rely on correlated signatures. The two near-caustic image tracks merge along a characteristic parabolic relation in the time--frequency plane, with the curvature direction distinguishing the two Gaussian-lens caustics. The third image has a predictable brightness and arrival-time ordering, while the same ray separation that produces differential dispersion may generate polarization changes in a sufficiently magnetized lens. For repeaters, transverse motion also predicts a chromatic migration of the lensing event between observing bands. Measurements of the delay, magnification, envelope width, fringe spacing, image multiplicity, differential DM or RM, and temporal evolution can therefore overconstrain the lens model.

The present analysis is restricted to a one-dimensional Gaussian profile, chosen because it makes these correlations transparent. Two-dimensional and non-Gaussian lenses may produce additional caustic morphologies, image configurations, and degeneracies, and will be considered in a separate work. We also examined whether isolated strong lenses can arise in volume-filling turbulent screens. Clean lensing occupies only a narrow transition between rare caustic-producing fluctuations and a confused network of overlapping caustics. This fine tuning suggests that volume-filling turbulence is not the most promising setting for identifiable lensing events. Sparse, narrow structures such as sheets, filaments, interfaces, or shocked clumps are more favorable because they can provide large column-density curvature without producing many competing images.

The most useful observational strategy is to search for several mutually consistent signatures rather than for isolated narrow-band bursts or apparently repeated morphologies. Finely channelized, high-time-resolution, full-Stokes voltage data are especially valuable for resolving image tracks and fringes and for measuring their polarization dependence. Such observations could turn plasma lensing from a flexible explanation of individual FRB features into a falsifiable probe of lens structure, FRB source size, and the fidelity with which multiple propagation paths preserve the emitted electric field. Most importantly, once a candidate lensing feature fixes a subset of the lens parameters, the remaining observables in Table~\ref{tab:lensing_observables} become
predictions rather than adjustable features.

\section*{Acknowledgments}
We acknowledge support provided by the NASA grant 80NSSC24K0770 for this work.
PB's work was also funded by a grant (no. 2024788) from the United States-Israel Binational Science Foundation (BSF), Jerusalem, Israel and by a grant (no. 1649/23) from the Israel Science Foundation.

\appendix
\restartappendixnumbering

\section{Spectrum of a lensed source}
\label{sec:spectrum}

\newcommand{\lowersub}[2]{%
  _{\raisebox{-#1}{$\scriptstyle #2$}}%
}
\newcommand{\uinu}{u_{\scriptscriptstyle I\nu_0}}

We consider the source and observer locations to be fixed. Consider that at some $\nu_0$ the magnification of the source is $\mu(\nu_0)$, and the image is located at $\uinu$. We wish to calculate $\mu(\nu)$ for $\nu$ near $\nu_0$. Since $\alpha\,\propto\, \nu^{-2}$, the image location varies with $\nu$, 
\begin{equation}
\label{eq:uIofnu}
 u_{\rm I}(\nu)=\uinu+\epsilon(\nu).   
\end{equation}
Expanding the image magnification expression  (Eq. \ref{eq:muGausslens}),  $\mu(u_{\rm I})^{-1}\!=\!1+\alpha(1-2u_{\rm I}^2)\exp(-u_{\rm I}^2)$, up to linear order in $\epsilon(\nu)$ we obtain
\begin{eqnarray}
  &  \mu^{-1}(\nu)\approx 1+\alpha(\nu)[1-2\uinu^2-4\epsilon\uinu]e^{-\uinu^2-2\epsilon\uinu} 
  \approx1+\alpha(\nu)[1-2\uinu^2-6\uinu\epsilon + 4\uinu^3\epsilon]e^{-\uinu^2}. 
\end{eqnarray}
Defining $\Delta \alpha=\alpha(\nu)-\alpha(\nu_{\rm 0})$ and using the magnification at $\nu_{\rm 0}$ to substitute $e^{-\uinu^2}=\frac{\mu^{-1}(\nu_0)-1}{\alpha(\nu_0)(1-2\uinu^2)}$ we get
\begin{eqnarray}
\label{eq:Deltamuofnu}
  &  \mu^{-1}(\nu)-\mu^{-1}(\nu_0) \approx \frac{\Delta \alpha [\mu^{-1}(\nu_0)-1]}{\alpha(\nu_0)} -\frac{2\alpha(\nu)\uinu\epsilon(3-2\uinu^2)(\mu^{-1}(\nu_0)-1)}{\alpha(\nu_0)(1-2\uinu^2)} 
\end{eqnarray}

To complete the process we need to determine $\epsilon(\nu)$. Using Eq. \ref{eq:uIx}, we write the expression for $\epsilon$:
\begin{eqnarray}
\label{eq:spectrumdenom}
 &  (\uinu \!+\! \epsilon) \left[1\!+\! \alpha(\nu)e^{-\uinu^2\!-\!2\epsilon\, u_{_{\rm I\nu_0}}}\right]\!=\!u_{s}\!=\!\uinu[1\!+\!\alpha(\nu_0)e^{-\uinu^2}] \implies \epsilon [\alpha(\nu)\mu^{-1}(\nu_0)-\Delta \alpha]=\frac{\uinu\Delta \alpha (1-\mu^{-1}(\nu_0))}{1-2\uinu^2}.
\end{eqnarray}
Plugging this back into Eq. \ref{eq:Deltamuofnu}, we get
\begin{eqnarray}
\label{eq:Delmuofnufin}
    & \mu^{-1}(\nu)-\mu^{-1}(\nu_0)= \frac{\Delta \alpha (\mu^{-1}(\nu_0)-1)}{\alpha(\nu_0)} + \frac{2\alpha(\nu)\uinu^2(3-2\uinu^2)[\mu^{-1}(\nu_0)-1]^2\Delta\alpha}{\alpha(\nu_0)(1-2\uinu^2)^2[\alpha(\nu)\mu^{-1}(\nu_0)-\Delta \alpha]}. 
\end{eqnarray}
Eq. \ref{eq:Delmuofnufin} gives $\mu(\nu)$. The observed spectrum is $f_{\rm obs}(\nu)=\mu(\nu) f(\nu)$, where $f(\nu)$ is the intrinsic FRB spectrum. For an unresolved source, such as an FRB, as long as the time delay between images is shorter than the burst duration, we need to add the contributions from different images. Whether these add up coherently (electric field summation) or incoherently (magnification summation), depends on the mutual coherence of the two image fields, as discussed in \S \ref{sec:images_interact}.
The magnification spectrum is shown in Fig. \ref{fig:spectrum}, alongside a comparison between the numerical and analytic treatments. The approximation given by Eq. \ref{eq:Delmuofnufin} works well except when one of the images passes very close to a caustic (which happens for the negatively magnified image at some $\nu>\nu_0$).
In particular, this expansion assumes that $\epsilon$ remains small and that the denominator in Eq. \ref{eq:spectrumdenom} does not approach zero; the latter corresponds to the image encountering a caustic at the shifted frequency.

\section{Internal scattering within an individual lens}
\label{app:internal_scattering}
An individual plasma lens may have structure on scales much smaller than its overall transverse size. Internal density fluctuations produce angular scattering in addition to the coherent deflection responsible for the macroscopic images. Let $\ell_{\phi,\rm int}$ denote the internal diffractive scale, defined such that the rms phase difference across this separation is of order unity. The corresponding scattering angle and temporal
broadening are approximately
\begin{equation}
 \theta_{\rm sc,int}\sim\frac{\lambda}{2\pi\ell_{\phi,\rm int}}, \qquad\tau_{\rm sc,int}\sim\frac{\lambda^2d_{\rm eff}}{8\pi^2c\,\ell_{\phi,\rm int}^2}.
 \label{eq:internal_scattering}
\end{equation}

Two macroscopic burst images remain distinguishable only if their scatter broadening does not exceed their group-delay separation,
\begin{equation}
 \tau_{\rm sc,int}\lesssim|\Delta t_g|,\qquad\ell_{\phi,\rm int}\gtrsim\left(\frac{\lambda^2d_{\rm eff}}{8\pi^2c|\Delta t_g|}\right)^{1/2}.
 \label{eq:internal_scattering_copies}
\end{equation}
Using Eq.~\ref{eq:near_caustic_group_delay}, this gives
\begin{eqnarray}
\label{eq:internal_diffractive_limit}
& \frac{\ell_{\phi,\rm int}}{a}\!\gtrsim \!\frac{\lambda d_{\rm eff}}{4\pi a^2}\begin{cases}  \langle\mu_{\rm tot}\rangle^{1/2}, & {\rm caustic~A},\\[0pt]\left[\langle\mu_{\rm tot}\rangle \mathcal{L}_{\alpha}\right]^{1/2}, & {\rm caustic~B}, \end{cases}  \sim\! \begin{cases} 7.5\times10^{-7}\, \frac{d_{{\rm eff},19}}{a_{13}^{2}\nu_9} \left(\dfrac{\langle\mu_{\rm tot}\rangle}{10}\right)^{1\over 2}, & {\rm caustic~A},\\[7pt] 1.7\times10^{-6}\, \frac{d_{{\rm eff},19}}{a_{13}^{2}\nu_9} \left(\dfrac{\langle\mu_{\rm tot}\rangle}{10}\right)^{1 \over 2} \left[ \dfrac{\mathcal{L}_{\alpha}} {5}\right]^{1\over 2}, & {\rm caustic~B}.
 \end{cases}  
\end{eqnarray}
Thus, for the canonical parameters, the internal diffractive scale may be approximately six orders of magnitude smaller than the lens size without erasing resolved burst copies.

Preserving a regular fringe pattern is more demanding. Since the internal scattering transfer function varies over a decorrelation
bandwidth $\Delta\nu_{\rm sc,int}\sim\tau_{\rm sc,int}^{-1}$, it should vary little across one lensing fringe:
\begin{equation}
 \Delta\nu_{\rm sc,int}\gg\Delta\nu_{\rm fr}, \qquad\hbox{or}\qquad \tau_{\rm sc,int}\ll|\Delta t_g|.
\end{equation}
In addition, the scattering transfer functions sampled by the two macroscopic images must remain sufficiently correlated. Internal scattering can therefore preserve distinguishable burst copies while still reducing or irregularizing their interference fringes.

\section{Partial coherence and visibility}
\label{visibility}

The interference of the two image fields is governed by their first-order mutual coherence after propagation and spectral filtering. For two faithful delayed copies of the same emitted field, this may be expressed in terms of the temporal coherence function associated with the spectrum transmitted to the measurement, which is defined as
\begin{equation}
g^{(1)}(t,\tau)\equiv \frac{\langle E(t)E^*(t-\tau)\rangle} {\left[  \langle|E(t)|^2\rangle  \langle|E(t-\tau)|^2\rangle \right]^{1/2}}.
\end{equation}
where $E(t)$ denotes the complex analytic electric field in a sufficiently narrow frequency interval and the averages are local compared with the evolution of the burst envelope. We suppress the explicit dependence on $t$ and observing frequency below.
The mutual coherence of the two lensed images is therefore $g^{(1)}(\Delta t_g)$. 

We distinguish between the intrinsic temporal coherence of the emitted radiation, $g_{\rm int}^{(1)}(t,\tau)$ and the mutual coherence of the two image fields in the measurement, $g^{(1)}(t,\tau)$ (the former is given by the same expression as the latter but with the observed fields replaced by the emitted ones). We may operationally define an intrinsic coherence time $t_{\rm coh}$ by, e.g., 
$|g_{\rm int}^{(1)}(t_{\rm coh})|=e^{-1}$. 

Writing $g^{(1)}(\tau)=|g^{(1)}(\tau)|\exp[i\phi_g(\tau)]$, the interference term in Eq. \ref{eq:mu_partial_coh} is
\begin{equation}
    2|\mu_+\mu_-|^{1/2}|g^{(1)}(\Delta t_g)| \cos\left[\Delta\Phi+\Delta\phi_{\rm M} +\phi_g(\Delta t_g)\right].
    \label{eq:mu_partial_coh_cos}
\end{equation}
Thus partial temporal coherence reduces the fringe amplitude and may also introduce a phase offset. Ignoring instrumental averaging and finite-source effects, the corresponding spectral fringe visibility is
\begin{equation}
  {\cal V}_{\rm temp}(\nu)\equiv \frac{\mu_{\max}-\mu_{\min}}  {\mu_{\max}+\mu_{\min}}=\frac{2|\mu_+\mu_-|^{1/2}}{|\mu_+|+|\mu_-|} \left|g^{(1)}[\Delta t_g(\nu)]\right|.
\label{eq:visibility_g1}
\end{equation}
Near a fold, $|\mu_+|\simeq|\mu_-|$, and hence 
\begin{equation}
{\cal V}_{\rm temp}\simeq \left|g^{(1)}(\Delta t_g)\right|.   
\label{eq:visibility_equal_images}
\end{equation}
More generally,
\begin{equation}
\label{eq:visibility}
{\cal V}_{\rm obs}\simeq {\cal V}_{\mu}|g^{(1)}(\Delta t_g)|{\cal V}_{\rm src}{\cal V}_{\rm ch}{\cal V}_{\rm pol}{\cal V}_{\rm sc},
\end{equation}
where ${\cal V}_{\mu}$ is the reduction caused by unequal image magnifications, ${\cal V}_{\rm src}$ accounts for averaging over the emitting region, ${\cal V}_{\rm ch}$ for finite spectral resolution, ${\cal V}_{\rm pol}$ for incomplete overlap of the image polarization states and ${\cal V}_{\rm sc}$ for scattering.
For a stationary noise-like field, sufficiently fine channelization of two faithful delayed copies ensures mutual coherence at the image delay; thus high visibility is expected for faithful delayed copies, whereas a significant deficit is the more informative result \footnote{A channel of width $\Delta\nu_{\rm ch}$ has a coherence time $\sim\Delta\nu_{\rm ch}^{-1}$; for a rectangular bandpass, $|g_{\rm ch}^{(1)}(\tau)|= |\mathrm{sinc}(\pi\Delta\nu_{\rm ch}\tau)|$. Resolving the lensing fringes requires $\Delta\nu_{\rm ch}\ll\Delta\nu_{\rm fr}\simeq|\Delta t_g|^{-1}$, and therefore $|g_{\rm ch}^{(1)}(\Delta t_g)|\simeq1$. Reduced visibility then indicates source-size averaging, polarization mismatch, differential propagation, or other failure of the images to remain faithful copies.}.
A lensed image pair can thus act as a temporal interferometer. After accounting for the observed spectrum, instrumental bandpass, unequal image magnifications, finite source size, polarization mismatch, and additional propagation effects, the fringe contrast constrains the mutual coherence of the two image fields at the image delay.

For a Gaussian source with projected rms size $\sigma_x$, the corresponding visibility factor is
\begin{equation}
\label{eq:visibilisrc}
{\cal V}_{\rm src}=\exp\left[-\frac12\left(\frac{8\pi a\sigma_x} {R_F^2|\eta|\langle\mu_{\rm tot}\rangle} \right)^2\right].
\end{equation}
A high fringe contrast requires both $|g^{(1)}(\Delta t_g)|$ and ${\cal V}_{\rm src}$ to be large and yields a lower bound on the former and an upper bound on the source size.

\end{document}